\documentclass[draft,tightenlines,nofootinbib,preprint,aps,eqsecnum,amsmath,amssymb]{revtex4}

\begin{document}

\newcommand{\beq}{\begin{equation}}
\newcommand{\eeq}{\end{equation}}
\newcommand{\bea}{\begin{eqnarray}}
\newcommand{\eea}{\end{eqnarray}}
\newcommand{\cir}{{\buildrel \circ \over =}}

\newcommand{\on}{\stackrel{\circ}{=}}
\newcommand{\byd}{\stackrel{def}{=}}
\baselineskip 20pt

\title{Quantum Mechanics in Non-Inertial Frames with a
Multi-Temporal Quantization Scheme: II) Non-Relativistic
Particles.}

\author{David Alba}

\affiliation
{Dipartimento di Fisica\\
Universita' di Firenze\\
Via G. Sansone 1\\
50019 Sesto Fiorentino (FI), Italy\\
E-mail: ALBA@FI.INFN.IT}

\begin{abstract}

The non-relativistic version of the multi-temporal quantization
scheme of relativistic particles in a family of non-inertial
frames (see hep-th/0502194) is defined. At the classical level the
description of a family of non-rigid non-inertial frames,
containing the standard rigidly linear accelerated and rotating
ones, is given in the framework of parametrized Galilei theories.
Then the multi-temporal quantization, in which the gauge
variables, describing the non-inertial effects, are not quantized
but considered as $c$-number generalized times, is applied to non
relativistic particles. It is shown that with a suitable ordering
there is unitary evolution in all times and that, after the
separation of the center of mass, it is still possible to identify
the inertial bound states. The few existing results of
quantization in rigid non-inertial frames are recovered as special
cases.

\medskip

\today

\end{abstract}

\maketitle

\newpage

\section{Introduction}

In a preceding paper (referred to as {\bf I}) \cite{1}  a
relativistic quantum mechanics for a system of N positive-energy
particles in a family of relativistic non-inertial frames was
defined in the framework of {\em parametrized Minkowski theories}.

In this paper there is the  study of the non-relativistic limit of
this non-inertial quantum mechanics. A classical {\em parametrized
Galilei theory} is constructed where a choice of non-inertial
coordinates is realized as a gauge choice. This is done without
any use of the relativistic theory of reference {\bf I}, so that
knowledge of results of {\bf I} is not necessary. However it is
also shown that the non-relativistic parametrized theory can be
obtained by the relativistic one making the exact limit
$c\rightarrow\infty$. In this approach a very general notion of
non-inertial coordinates is used. Indeed, treating the
3-dimensional Newtonian Space as a flat 3-dimensional manifold,
non inertial coordinates are defined applying a {\em time
dependent coordinate transformation} on the inertial ones. The
non-inertial coordinates associated to the traditional accelerated
or rotating frames are found as particular cases of linear, {\em
rigid}, time-dependent coordinate transformations.

The corresponding quantum theory is obtained by means of the {\em
multi-temporal quantization scheme} for first class constraints,
in which only the particle degrees of freedom are quantized. The
gauge variables describing the frame-dependent inertial effects
are not quantized, but considered as c-number generalized times.
By means of a suitable ordering  as many coupled Schroedinger-like
equations as first class constraints are obtained satisfying the
same algebra as the Poisson bracket algebra of the classical
constraints. It is possible to define a Hilbert space, whose wave
functions depend on time {\it and} on the generalized times as
parameters. All the Hamiltonians in the Schroedinger-like
equations are self-adjoint operators and the scalar product is
independent from all the times, so that the evolution is unitary.
By choosing a path in the parametric space of the generalized
times (namely a non-inertial frame) the non-inertial,
self-adjoint, Hamiltonian for the non-inertial evolution can be
found. In the particular case of rigidly linear accelerated or
rotating frames known results are reproduced as special cases.
Moreover, it is possible to study the separation of the center of
mass from the relative variables and to apply it to the definition
of bound states in non-inertial frames.

\bigskip

In Section II a classical non-relativistic parametrized theory
with first class constraint is constructed and it is also shown
that the gauge fixing on the gauge variables are interpretable as
a choice of non-inertial, in general non-rigid, coordinates. In
Section III there is the study of the multi-temporal quantization
scheme, of the non-relativistic coupled Schroedinger-like
equations and of a scalar product  independent from all the times
so that the evolution is unitary. A path in the parametric space
of the generalized times allows to find the self-adjoint,
Hamiltonian for the non-inertial evolution. In Section IV, there
is the restriction of the general theory to the particular case of
the traditional non-inertial rigid frames and it is shown that the
previous attempts \cite{Klink,Plebanski,Greenberger} to define
non-inertial quantum mechanics in rigid non-inertial frames are
special sub-cases of the multi-temporal approach. In Section V, a
study of the separation between center of mass and relative
coordinates is done and some observation on the possible
definitions of bound states in non inertial frames both in the
rigid and  non-rigid cases. In Section VI there are some
concluding remarks. In Appendix A the equivalence between the
non-relativistic limit of relativistic theory of {\bf I} and the
non-relativistic parametrized theory is shown. In Appendix B there
is the study of the spinning particles case. In Appendix C Galilei
transformation and Galilei relativity are discussed.\newpage

\section{Classical Parametrized Galilei Theories}

\subsection{Some Geometric Definitions}

In a non-relativistic theory, each event is identified by its
position in a 3-dimensional {\em space} ${\bf R}^3$ and by its
{\em (absolute) time} $t$. In the 3-dimensional space there exist
a family of preferred reference frames: the {\em inertial
reference frames}. We choose one of these ones taking a basis of
unit vectors $(\hat{\i}_1,\hat{\i}_2,\hat{\i}_3)$ defining three
ortoghonal directions placed on a arbitrary (fixed) origin: the
position in space of each event will be identified by the its
cartesian coordinates $x^a$ on the cartesian axis so defined.
These coordinates are the {\em inertial coordinates}.

In some cases, it can be useful to use a four dimensional
Newtonian space-time ${\bf R}\times{\bf R}^3$ where the {\em
(absolute) time} $t$ and the inertial cartesian spatial
coordinates $\vec{x}$ represent a particular choice of
4-dimensional coordinates. From a mathematical point of view, the
Newtonian space-time can be regarded as a 4-dimensional manifold
where a more general atlas of coordinates can be used such that,
for each chart, $\xi^\mu=\xi^\mu(t,\vec{x})$. The Cartan's
approach to Newtonian space-time \cite{Cartan}, used for example
in Ref.\cite{Kuchar}, is based on this observation. 

Here,
following Ref.\cite{Havas}, we use a more simple approach where we
admit time dependent coordinates transformations only on the
3-dimensional space. After one of these transformations the
3-dimensional space is parametrized by a set of global, in general
non cartesian coordinates $\sigma^r$. This {\em invertible, global
coordinates transformation} has the form ($a,r =1,2,3$)

\begin{equation}
x^a={\cal A}^a(t,\vec{\sigma}),
 \label{parametri}
\end{equation}

\noindent with inverse

\begin{equation}
\sigma^r=S^r(t,\vec{x}).
 \end{equation}

\noindent If we define the 3-dimesional Jacobian

\begin{equation}
J^a{}_r(t,\vec{\sigma})=\frac{\partial {\cal A}^a(t,\vec{\sigma})}{\partial\sigma^r},
\end{equation}

\noindent the invertibility conditions are

\begin{equation}
\det J(t,\vec{\sigma})>0.
\end{equation}

\noindent In particular we will use the inverse Jacobian

\begin{equation}
\widetilde{J}^r{}_a(t,\vec{\sigma})=\left[ \frac{\partial
S^r(t,\vec{x})}{\partial x^a}\right]_{\vec{x}
=\vec{S}(t,\vec{\sigma})},
\end{equation}

\noindent satisfying

\begin{equation}
J^a{}_r(t,\vec{\sigma})\,\widetilde{J}^r{}_b(t,\vec{\sigma})=\delta^a_b,
\;\;\;\;\;\;\;
\widetilde{J}^s{}_a(t,\vec{\sigma})\,J^a{}_r(t,\vec{\sigma})=\delta^s_r.
\end{equation}

\bigskip

The $\sigma^r$'s are the {\em non-inertial coordinates} of a
{non-inertial reference frame} implicitly defined by
Eq.(\ref{parametri}). As we shall see in Section IV, the
traditional linear accelerated and rotating non-inertial frames
(the {\em rigid frames}  obtained by means of rigid time dependent
transformations), are particular cases of this more general
approach.

\bigskip

In the following subsections we develop a Lagrangian and
Hamiltonian theory where the choice of  the non-inertial
coordinates $\sigma^r$, that is the choice of the coordinate
transformation (\ref{parametri}) can be interpreted as a
gauge-choice. This is done by introducing a singular Lagrangian
theory where the gauge parameters, namely the functions ${\cal
A}^a(t,\vec{\sigma})$, are considered as Lagrangian configuration
variables. At the Hamiltonian level this implies the presence of
first class Dirac constraints.

\subsection{Lagrangian Formulation}

If we apply the coordinates transformation (\ref{parametri}) to a
system of $N$ non-relativistic interacting particles, the particle
coordinates $x^a_i(t)$, $i=1,..,N$, in the inertial frame must be
parametrized in the form

\begin{equation}
x^a_i(t)={\cal A}^a(t,\vec{\eta}_i(t)),
 \label{posizione}
\end{equation}

\noindent so that the standard velocities acquire the following
form

\begin{eqnarray}
\dot{x}^a_i(t)&=&\frac{d{\cal A}^a(t,\vec{\eta}_i(t))}{dt}=\nonumber\\
&&\nonumber\\
&\byd&\left[ \frac{\partial {\cal
A}^a(t,\vec{\eta}_i(t))}{\partial t}
+J^a{}_r(t,\vec{\eta}_i(t))\,\frac{d\eta^r_i(t)}{dt}\, \right].
 \label{velocita}
\end{eqnarray}
If we apply these transformations to the usual inertial equations
of motion \footnote{ We use an interaction potential ${\bf
V}(t,\vec{x}_1,...,\vec{x}_N)$ which is time-dependence and
without rotational or translational symmetries. This is useful for
the discussion of the examples of Section VI. As shown in Appendix
C, the equations of motion are invariant under Galilei
transformations only if the interaction potential is time
independent and invariant under rotations and translations. }

\begin{equation}
m_i\, \frac{d}{dt}\, {\dot
x}^a_i(t) = -\frac{\partial {\bf V}}{\partial
x^a_i}\Big(t,\vec{x}_1(t),...,\vec{x}_N(t)\Big),
\end{equation}

\noindent we obtain

\begin{equation}
m_i\,\, \frac{d}{dt} \;\frac{d{\cal A}^a(t,\vec{\eta}_i(t))}{dt}=
- \left.\frac{\partial {\bf V}}{\partial x^a_i}\right|_{{\vec x}_i={\vec
{\cal A}}(t,{\vec \eta}_i(t))}
 = -\widetilde{J}^r{}_a(t,\vec{\eta}_i(t))\sum_j\frac{\partial
{\bf V}}{\partial\eta_i^r}.
 \label{EL_particelle}
\end{equation}

\noindent From now on ${\bf V}$ will denote the following expression for the
potential

\begin{equation}
{\bf V}={\bf V}\Big(t, \vec{\cal
A}(t,\vec{\eta}_1(t)),...,\vec{\cal A}(t,\vec{\eta}_N(t)) \Big).
 \label{potential}
\end{equation}

\bigskip

Let us consider the following action

\begin{equation}
S=\int dt\,L(t),
\end{equation}

\noindent with the Lagrangian

\begin{equation}
L(t) = \int d^3\sigma\, \sum_i\, \delta^3(\vec \sigma - {\vec
\eta}_i(t))\, \frac{1}{2m_i}\, \left( {{\partial {\vec {\cal
A}}(t,\vec \sigma )}\over {\partial t}} + J^a{}_r(t,\vec \sigma
)\, {{d \eta^r_i(t)}\over {dt}} \right)^2 - {\bf V}.
 \label{lagrangiana}
\end{equation}

\noindent In the Lagrangian (\ref{lagrangiana}) the configuration variables
are the particles positions $\vec{\eta}_i(t)$ and the functions
${\cal A}^a(t,\vec{\sigma})$ (with velocities $\partial{\cal
A}^a(t,\vec{\sigma})/\partial t$). The Euler-Lagrange equations

\beq 
{{\delta S}\over {\delta {\vec \eta}_i(t)}} = 0, 
\eeq

\noindent generated by a
variation of the variables ${\vec \eta}_i(t)$, are just
Eqs.(\ref{EL_particelle}).

However under the local Noether transformations 

\bea
&&\delta \eta^a_i(t) =
\varepsilon^a(t,{\vec \eta}_i(t))\, \widetilde{J}^r{}_a(t,{\vec \eta}_i(t)),\nonumber\\
&&\nonumber\\
&&\delta {\cal A}^a(t,\vec \sigma ) =\varepsilon^a(t,\vec \sigma ),
\eea

\noindent depending upon the arbitrary functions $\varepsilon^a(t,\vec \sigma )$, the
Lagrangian is invariant $\delta L = 0$. This implies that the
Euler-Lagrange equations for ${\cal A}^a(t, \vec{\sigma})$, are
not independent from the Eqs.(\ref{EL_particelle}), but satisfy
the contracted Bianchi identities 

\beq
{{\delta S}\over {\delta {\vec
{\cal A}}(t, \vec \sigma )}} = \sum_i\, \delta^3(\vec \sigma -
{\vec \eta}_i(t))\, \widetilde{J}^r{}_a(t, {\vec \eta}_i(t))\,
{{\delta S}\over {\delta \eta^r_i(t)}} = 0. 
\eeq

\noindent This means that the
configuration variables ${\cal A}^a(t,\vec{\sigma})$ are left
arbitrary (in other terms they are {\em gauge variables}) and that
at the Hamiltonian level there are {\em first class Dirac
constraints in the phase space}.

\bigskip

The Lagrangian (\ref{lagrangiana}) defines a {\it parametrized
Galilei theory}. See Appendix C for the study of the Galilei
transformations.

\subsection{Hamiltonian Formulation}

From the Lagrangian (\ref{lagrangiana}) we get the following
canonical {\em momenta}

\begin{eqnarray}
p_{ir}(t)&=&\frac{\partial L}{\partial \left(\frac{d\,
\eta^r_i(t)}{d t}\right)}= \,m_i
\,\sum_a\,J^a{}_r(t,\vec{\eta}_i(t))\, \frac{d{\cal
A}^a(t,\vec{\eta}_i(t))}{dt},\nonumber\\
&&\nonumber\\
\rho^a(t,\vec{\sigma})&=&\frac{\delta L}{\delta
\left(\frac{\partial {\cal A}^a(t,\vec{\sigma})}{\partial
t}\right)}= \sum_i\,m_i\, \frac{d{\cal
A}^a(t,\vec{\eta}_i(t))}{dt} \delta^3(\vec{\sigma}-\vec{\eta}_i),
 \label{momenti}
\end{eqnarray}

\noindent whose  Poisson brackets are

\begin{eqnarray}
&&\{\eta^r_i(t),p_{js}(t)\}=\delta^r_s\,\delta_{ij}\nonumber\\
&&\nonumber\\
&&\{{\cal A}^a(t,\vec{\sigma}),
\rho^b(t,\vec{\sigma}^{\,\prime})\} = \,\delta^{ab}\,
\delta^3(\vec{\sigma}-\vec{\sigma}^{\,\prime}).
 \label{PB}
\end{eqnarray}

\noindent The momenta satisfy the following Dirac constraints

\begin{equation}
{\cal H}_a(t,\vec{\sigma})= \rho^a(t,\vec{\sigma}) - \sum_i\,
\delta^3(\vec{\sigma}-\eta_i(t))\,\widetilde{J}^r{}_a(t,\vec{\eta}_i(t))\,p_{ir}(t)
\approx 0.
 \label{H_a}
\end{equation}

\noindent The canonical Hamiltonian is

\begin{eqnarray}
H_c(t)&=&
+\sum_i\,p_{ir}(t)\frac{\partial\eta^r_i(t)}{\partial t}+
\int d^3\sigma\,\sum_a\,\rho^a(t,\sigma)\,
{\cal A}^a(t,\vec{\sigma})-L(t)=\nonumber\\
&&\nonumber\\
&=&\sum_i\,\frac{1}{2m_i}\sum_a\,
\left[\widetilde{J}^r{}_a(t,\vec{\eta}_i(t))\,
p_{ir}(t)\right]\,
\left[\widetilde{J}^s{}_a(t,\vec{\eta}_i(t))\,
p_{is}(t)\right]
+{\bf V},
\end{eqnarray}

\noindent but the motion equations are generated by the Dirac
Hamiltonian

\begin{equation}
H_D(t)=H_c(t)+ \int d^3\sigma\,\lambda^a(t,\vec{\sigma})\,{\cal
H}_a(t,\vec{\sigma}),
 \label{hamiltoniana_dirac}
\end{equation}

\noindent where $\lambda^a(t,\vec{\sigma})$ are arbitrary Dirac's
multipliers. The constraints (\ref{H_a}) are first class Dirac
constraints because their  Poisson brackets are

\begin{equation}
\{H_c(t), {\cal H}_a(t,\vec{\sigma})\} = \{{\cal
H}_a(t,\vec{\sigma}), {\cal H}_b(t,\vec{\sigma}^{\,\prime})\}=0.
\end{equation}

\noindent These constraints are the canonical generators of the Noether
gauge transformations, corresponding to a change in the definition
of the non inertial coordinates $\vec{\sigma}$, under which the
Lagrangian is invariant.

\bigskip

The equations of motion generated by the Dirac Hamiltonian
(\ref{hamiltoniana_dirac}) imply

\begin{equation}
\lambda^a(t,\vec{\sigma})\on\frac{\partial {\cal
A}^a(t,\vec{\sigma})}{\partial t}.
\end{equation}

\noindent A gauge fixing for the first class constraints is a choice of a
{\em fixed} form of the function ${\cal A}^a(t,\vec{\sigma})$
\footnote{We use the subscript "F" for the geometrical quantities
calculated in a fixed gauge.}

\begin{equation}
{\cal A}^a(t,\vec{\sigma})={\cal A}^a_F(t,\vec{\sigma}),
\label{gauge-fixing}
\end{equation}

\noindent and this is equivalent to fix the arbitrary multipliers:

\begin{equation}
\lambda^a(t,\vec{\sigma})=\frac{\partial {\cal
A}_F^a(t,\vec{\sigma})}{\partial t}.
 \label{gauge-fixing2}
\end{equation}

\noindent After such a choice the only  remaining canonical variables are
$\eta^r_i(t)$, $p_{ir}(t)$. The equations of motion for a function
$F(t,\vec{\eta}_i,\vec{p}_i)$ of the particle canonical variables,
evaluated in the fixed gauge, are the Hamilton-Dirac equations
generated by the Hamiltonian (\ref{hamiltoniana_dirac}) and
restricted to Eqs. (\ref{gauge-fixing}) and (\ref{gauge-fixing2})

\begin{equation}
\left(\frac{d}{dt}F\right)_{{\cal A}={\cal A}_F}\on \left[
\frac{\partial}{\partial
t}F+\{F,H_c\}-\sum_i\{F,\lambda^a(t,\vec{\eta}_i)\,
\widetilde{J}^r{}_a(t,\vec{\eta}_i(t))\,p_{ir}\}\right]_{{\cal
A}={\cal A_F}}.
 \label{eq_moto1}
\end{equation}

Since the gauge fixings are explicitly time dependent, it can be
shown that in the non-inertial frame corresponding to the
given gauge these equations are generated by the non inertial
Hamiltonian

\begin{equation}
H_{ni}(t)=\sum_i\,\sum_a\,
\frac{\left[\widetilde{J}_F^r{}_a(t,\vec{\eta}_i(t))\,
p_{ir}(t)\right]\,
\left[\widetilde{J}_F^s{}_a(t,\vec{\eta}_i(t))\,
p_{is}(t)\right]}{2m_i}
+{\bf V}-\sum_i\,V^r_F(t,\vec{\eta}_i)\,p_{ir}(t),
\label{H-classica-ni}
\end{equation}

\noindent where we have introduced the velocity field

\begin{equation}
V^r_F(t,\vec{\sigma})= \widetilde{J}_F^r{}_a(t,\vec{\sigma})\,
\frac{\partial {\cal A}^a_F(t,\vec{\sigma})}{\partial t}.
\end{equation}

\noindent Then Eqs.(\ref{eq_moto1}) can be written in the form

\begin{equation}
\left(\frac{d}{dt}F\right)_{{\cal A}={\cal A}_F}\on
\frac{\partial}{\partial t}F+\{F,H_{ni}(t)\}.
\end{equation}

\newpage

\section{Multi-temporal Quantization}

In the traditional Dirac Quantization scheme of a classical theory
with first class contraints \cite{Henneaux}, all the canonical
variables are quantized ignoring the presence of the constraints
and a {\em non-physical Hilbert space} is constructed. After a
suitable choice of the ordering, the constraints are mapped in
operators and {\em physical states} are defined as the zero
eingenvectors of the quantum constraints. The main difficulty of
this approach is the definition of a {\em physical Hilbert space},
with a {\em physical scalar product}, since usually the quantum
constraints have zero in their continuum spectrum and then the
{\em physical states} do not define a subspace of the {\em
non-physical Hilbert Space}.

When it is possible to identify the {\em gauge variables}
associated to Dirac constraints (as it happens in the parametrized
theory of previous Section, where it is possible to identify the
${\cal A}^a(t,\vec{\sigma})$ as the {\em gauge variables}), we can
avoid the construction of the {\em non-physical Hilbert space}
defining directly the {\em physical Hilbert space}. Indeed,
motivated by a classical multi-temporal approach to constrained
dynamics \cite{classic_manytimes1}, \cite{classic_manytimes2} and
by the use of an analogous many-time quantization scheme for two
particles relativistic systems \cite{Longhi}, we treat in a
different way the {\em gauge variables} and their momenta with
respect to the other {\em physical} variables. These latter are
quantized and interpreted as operators and the physical Hilbert
space is constructed so to realize an irreducible representation
of their canonical commutation relations. Instead the gauge
variables are interpreted as {\em generalized times} and their
momenta are mapped in time derivatives. The contraints are mapped
in  coupled generalized Schroedinger equations that govern the
dependence on the generalized times of the wave function. One has
to find an ordering for the quantization of the constraints such
that the quantum algebra of the constraints implies the
integrability of the generalized Schroedinger equations. The
physical scalar product in the Hilbert space has to be independent
from all the generalized times.

This scheme of quantization has already been used in {\bf I} for
parametrized Minkowski theories.

\subsection{Quantization: "Times", Operators and Hilbert Space.}

We go now to apply the {\em multi-temporal quantization scheme} to
the  parametrized Galilei theory of the previous Section.

i) We shall consider the gauge variables ${\cal
A}^a(\vec \sigma )$ as {\it c-number generalized times}, with the
conjugate momenta replaced by the following functional derivatives

\beq
 \rho^a(\vec{\sigma})\mapsto \,-i\hbar\,
\frac{\delta}{\delta {\cal A}^a(\vec{\sigma})}.
\label{4B.1}
\eeq

ii) The  positions and  momenta $\eta^r_i,\kappa_{i\,r}$ of the
particles are quantized in the standard way as operators on a
Hilbert space satisfying the canonical commutation relations. We
choose a representation where the $\eta^r_i$'s are multiplicative
operators and where

 \beq
  p^r_i \mapsto -i\, \hbar\,
\frac{\partial}{\partial\eta^r_i},
 \label{4B.2}
  \eeq

\noindent are derivative operators on the {\em Hilbert space} of
square integrable complex functions ${\bf H}=L^2({\bf C},{\bf
R}^{3N})$ with scalar product \footnote{ In  I, it is shown that
we can develop the quantum theory using also a scalar product with
the {\em invariant measure}
\[
(\Phi_1,\Phi_2)_{inv}=\int \left( \prod_i\,\det
J(t,\vec{\eta}_i)\,d^3\eta_i\right)\,
\overline{\Phi}_1(\vec{\eta_i}) \Phi_2(\vec{\eta_i}).
\]
The wave functions $\Phi$ have the following relation with the
$\Psi$'s
\[
\Psi(\vec{\eta}_i;t,{\cal A}^a]=\sqrt{\prod_i\,\det
J(t,\vec{\eta}_i)} \,\Phi(\vec{\eta}_i;t,{\cal A}^a].
\]
As discussed in I, in this approach the conservation of
probability
\[
\frac{\partial}{\partial t}\, (\Phi_1,\Phi_2)_{inv}=
\frac{\delta}{\delta {\cal A}^a(\vec{\sigma})} (\Phi_1,\Phi_2)_{inv}=0,
\]
is implied if the new wave functions satisfy generalized
Schoredinger equations with non-self-adjoint Hamiltonians, which
are obtained with a {\em non symmetrized} ordering choice in the
Eqs.(3.8) and (3.10). }

\begin{equation}
(\Psi_1,\Psi_2)=\int \left(\prod_i\,d^3\eta_i\right)\,
\overline{\Psi}_1(\vec{\eta_i}) \Psi_2(\vec{\eta_i}).
 \label{4B.4}
\end{equation}

\subsection{"Generalized" Temporal Evolution}

A state will evolve in the Hilbert State ${\bf H}$ as a functional
of the {\em time} $t$ and of the {\em generalized times} ${\cal
A}^a(\vec{\sigma})$. The evolution in these {\em generalized
times} is determined by the quantization of the classical Dirac
contraints in the form \footnote{ We use the notation
$\Psi\left(\vec{\eta}_i;t,
 {\cal A}^a\right]$
to indicate that the wave function is a function
of the positions $\vec{\eta}_i$ and of the time $t$, but a functional
of the {\em generalized times} ${\cal A}^a(\vec{\sigma})$.
}

\begin{equation}
\widehat{\cal H}_a(\vec{\sigma})\cdot\Psi\left(\vec{\eta}_i;t,
 {\cal A}^a\right]=0,
  \label{4B.12}
\end{equation}

\noindent whereas the evolution in the {\em time} $t$ is
determined by the quantum Dirac Hamiltonian through the
Schroedinger equation

\beq
 i\,\hbar\,\frac {\partial}{\partial t}\,
 \Psi\left(\vec{\eta}_i;t,
 {\cal A}^a\right] = \widehat{H}_c\cdot\Psi\left(\vec{\eta}_i;t,
 {\cal A}^a\right].
 \label{4B.11}
\eeq

\bigskip

The explicit form of Eqs.(\ref{4B.11}) and (\ref{4B.12}) is
obtained by using the rules i) and ii). To solve the ordering
problems, we define the operators

\begin{equation}
\widehat{K}_{ia}= - \frac{i\hbar}{2}\,
\left[\widetilde{J}^r{}_a(\vec{\eta}_i)\,,
\,\frac{\partial}{\partial\eta^r_i}\right]_+ =
-i\hbar\,\widetilde{J}^r{}_a(\vec{\eta}_i)\,\frac{\partial}{\partial\eta^r_i}
-\frac{i\hbar}{2}\,\frac{\partial
\widetilde{J}^r{}_a(\vec{\eta}_i)}{\partial\eta^r_i}.
\end{equation}

\noindent They are the self-adjoint observables corresponding to classical
functions $\widetilde{J}^r{}_a(\vec{\eta}_i)\,p_{ir}$. Then we
assume the following ordering inside the canonical Hamiltonian
$H_c$

\begin{equation}
\sum_a\,\widetilde{J}^r{}_a(\vec{\eta}_i)\,p_{ir}\;
\widetilde{J}^s{}_a(\vec{\eta}_i)\,p_{is}\longmapsto\,
-\hbar^2\Delta^{\,\prime}_i=\sum_a\widehat{K}_a\cdot\widehat{K}_a.
\end{equation}

\noindent Instead inside the constraints ${\cal H}_a(t,\vec{\sigma})$ we
introduce the ordering

\begin{equation}
\sum_i\delta^3(\vec{\sigma}-\vec{\eta}_i)\,\widetilde{J}^r{}_a(\vec{\eta}_i)\,p_{ir}
\longmapsto\,\widehat{T}_a(\vec{\sigma},{\cal
A}^a]=-\frac{i\hbar}{2}\sum_i\left[
\delta^3(\vec{\sigma}-\eta_i)\,\widetilde{J}^r{}_a(\vec{\eta}_i)\,
,\,\frac{\partial}{\partial\eta^r_i} \right]_+ .
\end{equation}

\noindent With these definition we obtain the {\em generalized Schroedinger
equations}:

\bea
 &&\left( i\, \hbar\, {{\partial}\over {\partial t}} - \widehat{E}[{\cal A}^a]\right)\,
 \Psi\left(\vec{\eta}_i;t,
 {\cal A}^a\right]
 = 0, \label{4B.18}\\
 &&{}\nonumber \\
 && \widehat{\cal H}_a(\vec{\sigma})\cdot\Psi\left(\vec{\eta}_i;t,
 {\cal A}^a\right]=
 \left(-i\, \hbar\,{{\delta}\over {\delta {\cal A}^s}(\vec{\sigma})}
 -\widehat{T}_a(\vec{\sigma},{\cal A}^a] \right)\,
 \Psi\left(\vec{\eta}_i;t,
 {\cal A}^a\right]= 0,
 \label{4B.19}
 \eea

\noindent where we used the notation

\begin{equation}
\widehat{H}_c\equiv\widehat{E}[{\cal A}^a]=
-\hbar^2\,\sum_i\,\frac{1}{2m_i}\,\Delta^{\,\prime}_i+{\bf V}.
\label{energia-ni}
\end{equation}

With this notation we want to emphasize that the operator
$\widehat{E}[{\cal A}^a]$ is the energy defined by the inertial
observer adopting the inertial coordinates $x^a$ expressed as
function of the coordinates $\vec{\sigma}$ defined by $x^a={\cal
A}^a(\vec{\sigma})$.

The chosen ordering  is such that the generalized Hamiltonians
$\widehat{E}[{\cal A}^a]$ and $\widehat{T}_a(\vec{\sigma},{\cal
A}^a]$ are self-adjoint operators and it {\em guarantees the
formal integrability} of Eqs.(\ref{4B.18}) and (\ref{4B.19}),
namely

\begin{equation}
\Big[\,\widehat{E}[{\cal A}^a],\widehat{\cal
H}_a(\vec{\sigma})\,\Big]= \Big[\,\widehat{\cal
H}_a(\vec{\sigma}),\widehat{\cal
H}_b(\vec{\sigma}^{\,\prime})\Big]=0.
\end{equation}

\bigskip

We can formalize the {\em generalized time evolution} by
introducing a {\em space of generalized times}, parametrized with
the {\em time} $t$ and with the {\em generalized times} ${\cal
A}^a(\vec{\sigma})$. Such a {\em space of generalized times} is
the cartesian product ${\cal M}={\bf R}\times C^\infty({\bf
R}^3,{\bf R}^3)$, where $C^\infty({\bf R}^3,{\bf R}^3)$ is the
space of the differentiable and invertible functions from ${\bf
R}^3$ on ${\bf R}^3$, whose elements are represented by $(t,{\cal
A}^a)$ [${\cal A}^a\in C^\infty({\bf R}^3,{\bf R}^3)$]. Then, the
{\em generalized temporal evolution} can be defined as the map

\beq
 {\cal T}:{\cal M}\times{\bf H} \mapsto{\bf H}, \label{4B.24}
\eeq \beq {\cal T}\Big[\, (t,{\cal A}^a),\Psi_o(\vec{\eta}_i)
\Big]=\Psi\left(\vec{\eta}_i,t,{\cal A}^a\right],
 \label{4B.25}
\eeq

\noindent where $\Psi_o(\vec{\eta}_i)$ is the {\it initial
condition}, namely the value assigned to the state in a point
$(t_o,{\cal A}^a_{in}(\vec{\sigma}))\in{\cal M}$

 \beq
\Psi\left(\vec{\eta}_i,t_o,{\cal
A}^a_{in}\right]=\Psi_o(\vec{\eta}_i).
 \label{condizione-iniziale}
\eeq

Since the generalized Hamiltonians are self adjoint we get

\begin{equation}
\frac{\partial}{\partial t}\, (\Psi_1,\Psi_2)=
\frac{\delta}{\delta {\cal A}^a(\vec{\sigma})} (\Psi_1,\Psi_2)=0.
 \label{4B.34}
\end{equation}

\noindent This implies that the {\em time evolution} ${\cal T}$ defines a
unitary transformation in Hilbert space {\bf H}.

\bigskip

To discuss this fact explicitly, it convenient to assign the {\em
initial condition} by choosing ${\cal
A}^a_{in}(\vec{\sigma})=\sigma^a$. Then the solution of the
generalized Schroedinger equations (\ref{4B.18}) and
(\ref{4B.19}), satisfying the initial condition
(\ref{condizione-iniziale}) when  evaluated at the generalized
times $t=t_o,{\cal A}^a(\vec{\sigma})= {\cal
A}^a_{in}(\vec{\sigma})$, is explicitly given by

\begin{eqnarray}
\Psi\left(\vec{\eta}_i,t , {\cal A}^a\right]&=&
\exp\left[-\frac{i}{\hbar}\,(t-t_o)\;
\widehat{E}[{\cal A}^a]\right]\cdot\Psi^{\,\prime}(\vec{\eta}_i;{\cal A}^a]=\nonumber\\
&&\nonumber\\
&=&\exp\left[-\frac{i}{\hbar}\,(t-t_o)\; \widehat{E}[{\cal
A}^a]\right]\cdot{\cal U}^{\,\prime}[{\cal A}^a]
\cdot\Psi_o(\vec{\eta}_i) \byd{\cal U}(t,{\cal
A}^a]\cdot\Psi_o(\vec{\eta}_i),
 \label{soluzione2}
\end{eqnarray}

\noindent where

 \beq
  \Psi^{\,\prime}(\vec{\eta}_i;{\cal
A}^a]=\sqrt{\prod_i\,\det J(\vec{\eta}_i)}\, \Psi_o\Big(\,
\vec{\cal A}(\vec{\eta}_i)\,\Big)\byd{\cal U}^{\,\prime}[{\cal
A}^a] \cdot\Psi_o(\vec{\eta}_i).
 \label{soluzione1}
  \eeq

\noindent If $\Psi^{\,\prime}_1,\Psi^{\,\prime}_2$ are obtained from two
different initial conditions $\Psi_{1,o},\Psi_{2,o}$ as in
Eq.(\ref{soluzione1}), it can be shown that, by using the change
of variables $\vec{\eta}^{\,\prime}_i=\vec{\cal A}(\vec{\eta}_i)$,
we get

 \beq
  \int \left(\prod_i\,d^3\eta^{\,\prime}_i\right)\,
\overline{\Psi}_{o,1}(\vec{\eta}^{\,\prime}_i)\,\Psi_{o,2}(\vec{\eta}^{\,\prime}_i)=
\int
\left(\prod_i\,d^3\eta_i\right)\,\overline{\Psi}^{\,\prime}_1(\vec{\eta}_i;{\cal
A}^a]\,\Psi^{\,\prime}_2(\vec{\eta}_i;{\cal A}^a].
 \label{4B.46}
 \eeq

\noindent Then Eq.(\ref{soluzione1}) defines a unitary transformation
on the Hilbert Space ${\bf H}$. The inverse of ${\cal
U}^{\,\prime}[{\cal A}^a]$ is also a self-adjoint operator

\begin{equation}
{\cal U}^{\,\prime\,+}[{\cal A}^a]
\cdot\Psi^{\,\prime}(\vec{\eta}^{\,\prime};{\cal A}^a]=
\frac{1}{\sqrt{\prod_i\,\det
J\Big(t,\vec{S}(\vec{\eta}^{\,\prime}_i)\Big)}}
\Psi^{\,\prime}\Big(\vec{S}(\vec{\eta}^{\,\prime}_i),{\cal
A}^a\Big].
 \end{equation}

\noindent Moreover, since $\exp\left[-\frac{i}{\hbar}\,(t-t_o)\;
\widehat{E}[{\cal A}^a]\right]$ is a unitary operator, also
Eq.(\ref{soluzione2}) defines a unitary transformation.

By using this result it can be shown that the values taken by a
solution (\ref{soluzione2}) in two different points $(t_1,{\cal
A}^a_1(\vec{\sigma}))$, $(t_2,{\cal A}^a_2(\vec{\sigma}))$, of the
space of the generalized times, are connected by a unitary
transformation

\begin{equation}
\Psi\left(\vec{\eta}_i,t_2 , {\cal A}^a_2\right]=
\exp\left[-\frac{i}{\hbar}\,(t_2-t_1)\; \widehat{E}[{\cal
A}^a]\right]\cdot {\cal U}^{\,\prime}[{\cal A}_2^a]\cdot {\cal
U}^{\,\prime\,+}[{\cal A}_1^a]\cdot \Psi\left(\vec{\eta}_i,t_1 ,
{\cal A}^a_1\right].
 \label{u}
\end{equation}

\subsection{Definition of a Non-Inertial Frame}

In the classical theory we select a non-inertial frame by fixing
the gauge variables $\vec{\cal A}(t,\vec{\sigma})$ as in
Eqs.(\ref{gauge-fixing}). At the quantum level, this can be
realized by defining a path

 \beq
  {\cal P}_F(t)=(t,{\cal
A}^a_F(t,\vec{\sigma})),
 \label{4B.53}
  \eeq

\noindent in the space of generalized times ${\cal M}$.

We assume the following point of view. Each physical state is
represented by the {\em generalized wave function}
(\ref{soluzione2}). The observer adopting a set of non-inertial
coordinates $\vec{\sigma}$, implicitly defined by the coordinates
transformation ${\cal A}^a_F(t,\vec{\sigma})$, will describe the
state by means of the wave function
$\Psi\left(\vec{\eta}_i,t,{\cal A}^a\right]$ evaluated along the
path ${\cal P}_F(t)$

\begin{equation}
\psi_F(t,\vec{\eta}_i)= \Psi\left(\vec{\eta_i},t;{\cal A}^a_F(t)\right].
 \label{4B.54}
\end{equation}

\noindent Since we have

\begin{eqnarray}
i\hbar\,\frac{\partial}{\partial t}\, \psi_F(t,\vec{\eta}_i)&=&
i\hbar\, \left[
\frac {\partial\Psi}
{\partial t}\right]\left(\vec{\eta_i},t ;{\cal A}_F
^a(t)\right] +\nonumber\\
 &&\nonumber\\
  &+&i\hbar\, \int d^3\sigma\,
\frac{\partial {\cal A}^a_F(t,\vec{\sigma})}{\partial t}\,
 \left[ \frac{\delta\Psi} {\delta {\cal A}
^a(\vec{\sigma})}\right] \left(\vec{\eta_i},t ;{\cal A}
^a_F(t))\right],
 \label{4B.55}
\end{eqnarray}

\noindent we see that Eqs.(\ref{4B.18}) and (\ref{4B.19}) {\em
imply the following non-inertial Schroedinger equation along the
path ${\cal P}_F(t)$ in the space of generalized times}

\bea
 i\hbar\,\frac{\partial}{\partial t}\, \psi_F(t,\vec{\eta}_i)&=&
\left[
\widehat{E}[{\cal A}_F^a]\, +i\hbar \sum_{i=1}^N\,\left(
V^r_F(t,\vec{\eta}_i)\, \frac{\partial}
{\partial\eta^r_i}+\frac{1}{2}\,
\frac{\partial V^r_F(t,\vec{\eta}_i)}{\partial \eta^r_i}\right)\right]\,
\psi_F(t,\vec{\eta}_i) =\nonumber \\
&&\nonumber\\
 &{\buildrel {def}\over =}& \widehat{H}_{ni}\cdot \psi_F(t ,{\vec
 \eta}_i).
 \label{H-quantistica-ni}
\end{eqnarray}

\noindent The non-inertial Hamiltonian operator $\widehat{H}_{ni}$ is just
the quantized version of the non-inertial Hamiltonian of
Eq.(\ref{H-classica-ni}). \hfill

For each value $t$, $\psi_F(t,\vec{\eta}_i)$ is a state in the
Hilbert space ${\bf H}$. The $t$-dependent non-inertial
Hamiltonian defined in Eq.(\ref{H-quantistica-ni}) is self adjoint
and the $t$-evolution along the path defines a unitary
transformation.

Indeed let $(t_1, {\cal A}^a_1(\vec{\sigma}) = {\cal
A}^a_F(t_1,\vec{\sigma}))$, $(t_2,{\cal A}^a_2(\vec{\sigma})={\cal
A}^a_F(t_2,\vec{\sigma}))$, be the extremal points of a path in
the generalized times space. Then Eq.(\ref{u}) implies that
$\psi_F(t_1,\vec{\eta}_i)$ and $\psi_F(t_2,\vec{\eta}_i)$ are
connected by a unitary transformations. Moreover this
transformation does not depend on the path but only on its
extremal points.

\bigskip

We want show that the non-inertial wave function
$\psi_F(t,\vec{\eta}_i)$ can be obtained by a time-dependent
unitary transformation from the wave function
$\psi_{in}(t,\vec{\eta}_i)$ defined in an inertial frame . To show
it, we choose another path ${\cal P}_{in}(t)=(t,{\cal
A}^a_{in}(t,\vec{\sigma}))$, where ${\cal A}^a_{in}(t,
\vec{\sigma})\equiv\sigma^a$. This path defines the inertial frame
with inertial coordinates $x^a=\sigma^a$. The wave function
restricted to this path is the wave function in the inertial
reference frame

\begin{equation}
\psi_{in}(t,\vec{\eta}_i)=\Psi(\vec{\eta}_i,t,{\cal A}_{in}^a].
\end{equation}

\noindent Moreover in this case Eq.(\ref{H-quantistica-ni}) becomes  the
usual inertial Schroedinger equation

\begin{eqnarray}
&&i\hbar\,\frac{\partial\psi_{in}}{\partial t}(t,\vec{\eta}_i)
=\widehat{E}_{in}\cdot\psi_{in}(t,\vec{\eta}_i),\nonumber\\
&&\nonumber\\
&&\widehat{E}_{in}=\widehat{E}[{\cal
A}^a_{in}]=-\hbar^2\,\sum_{i,a}
\frac{1}{2m_i}\,\left(\frac{\partial}{\partial\eta_i^a}\right)^2+
{\bf V}\Big(t,\vec{\eta}_1,...,\vec{\eta}_N\Big),
\label{energia-in}
\end{eqnarray}

\noindent and then we get

\begin{equation}
\psi_{in}(t,\vec{\eta}_i)=\exp\left[-\frac{i}{\hbar}\,(t-t_o)\,\widehat{E}_{in}\right]\cdot
\psi_{o,in}(t,\vec{\eta}_i).
 \label{evoluzione-inerziale}
\end{equation}

\noindent The two different observers using different sets of coordinates,
the inertial one defined by the path ${\cal P}_{in}(t)$ and the
non-inertial one defined by a path ${\cal P}_F(t)$, describe the
{\em same} physical system by evaluating the {\em same}
$\Psi(\vec{\eta}_i,T,{\cal A}^a]$ on {\em different} paths ${\cal
P}_{in}(t),{\cal P}_F(t)$, so that they obtain two {\em different}
wave functions $\psi_{in}(t,\vec{\eta}_i),\psi_F(t,\vec{\eta}_i)$
in ${\bf H}$. However, {\em there exist a time dependent unitary
transformation mapping $\psi_{in}$ to $\psi_{F}$}. To show this we
can use the general solution of Eqs.(\ref{soluzione1}) and
(\ref{soluzione2}) to write $\psi_F$ in terms of $\psi_{in}$.

1) In Eq.(\ref{condizione-iniziale}) let it be
$\Psi_o(\vec{\eta}_i)=\psi_{o,in}(\vec{\eta}_i)$. Then, using the
definitions (\ref{energia-ni}) and (\ref{energia-in}) and
Eq.(\ref{soluzione1}), it is only matter of tedious calculation to
show that, for each ${\cal A}^a(\vec{\sigma})$, we get

\begin{equation}
\widehat{E}[{\cal A}^a]\cdot\Psi^{\,\prime}(\vec{\eta}_i;{\cal
A}^a] =\sqrt{\prod_i\,\det J(\vec{\eta}_i)}\,\Big[
\widehat{E}_{in}\cdot\psi_{o,in}\Big]\Big(\vec{\cal
A}(\vec{\eta}_i)\Big)= {\cal U}^{\,\prime}[{\cal A}^a]\cdot\Big[\,
\widehat{E}_{in}\cdot\Psi_{o,in}(\vec{\eta}_i)\Big].
\label{proprieta-e}
\end{equation}

\noindent Then, using the result (\ref{proprieta-e}) and
Eq.(\ref{evoluzione-inerziale}), Eqs.(\ref{soluzione2}) become

\begin{equation}
\Psi(\vec{\eta}_i,t,{\cal A}^a]= \sqrt{\prod_i\,\det
J(\vec{\eta}_i)}\; \psi_{in}\Big(t,\vec{\cal
A}(\vec{\eta}_i)\Big).
\end{equation}

2) Finally, we can evaluate this general solution on the path ${\cal P}_F(t)$
 and we obtain the searched result

\begin{equation}
\psi_F(t,\vec{\eta}_i)= \sqrt{\prod_i\,\det J_F(t,\vec{\eta}_i)}\;
\psi_{in}\Big(t,\vec{\cal A}_F(t,\vec{\eta}_i)\Big) \byd {\cal
U}_F(t)\cdot\psi_{in}(t,\vec{\eta}_i). \label{trasf-unit-ni-in}
\end{equation}

\noindent This relation defines the time dependent unitary transformation
${\cal U}_F(t)$. The unitarity can be checked by making the change
of variables $\vec{\eta}^{\,\prime}_i=\vec{\cal
A}_F(t,\vec{\eta}_i)$ as in the transformation (\ref{soluzione1})

\begin{equation}
\int\left(\prod_i\,d^3\eta^{\,\prime}_i\right)\,
\overline{\psi}_{1,in}(t,\vec{\eta}^{\,\prime}_i)\,
\psi_{2,in}(t,\vec{\eta}^{\,\prime}_i) =\int
\left(\prod_i\,d^3\eta_i\right)\,
\overline{\psi}_{1,F}(t,\vec{\eta}_i)\,
\psi_{2,F}(t,\vec{\eta}_i).
\end{equation}

\bigskip

Using these results we can discuss Eq.(\ref{H-quantistica-ni})
from a new point of view. By using the form
(\ref{trasf-unit-ni-in}) for $\psi_F$ we can rewrite

\begin{equation}
\widehat{H}_{ni}(t)={\cal U}_F(t)\cdot\widehat{E}_{in}\cdot{\cal
U}^+_F(t) -{\cal U}_F(t)\frac {d{\cal U}^+_F(t)}{dt},
 \label{U1}
\end{equation}

\noindent where, as a consequence of Eq.(\ref{proprieta-e}),
$\widehat{E}[{\cal A}^a_F]= {\cal
U}_F(t)\cdot\widehat{E}_{in}\cdot{\cal U}^+_F(t)$ is {\em the
energy in a inertial reference frame written in non-inertial
coordinates} and where

\begin{equation}
-{\cal U}_F(t)\frac {d{\cal U}^+_F(t)}{dt} =+i\hbar
\sum_{i=1}^N\,\left( V^r_F(t,\vec{\eta}_i)\, \frac{\partial}
{\partial\eta^r_i}+\frac{1}{2}\,\frac{\partial
V^r_F(t,\vec{\eta}_i)}{\partial \eta^r_i}\right), \label{U2}
\end{equation}

\noindent is the {\em quantum inertial potential}.

\newpage

\section{Rigid Non-Inertial Frames}

In the previous Sections we worked with a very general definition
of non-inertial coordinates for non-relativistic mechanics.
Nevertheless, in a non-relativistic context rigidly linear
accelerated or rotating frames are usually used. The traditional
approach defines these non-inertial rigid frames by introducing a
time-dependent basis of unit vectors
$(\hat{b}_1(t),\hat{b}_2(t),\hat{b}_3(t))$ connected by a time
-dependent rotation to the fixed basis
$(\hat{\i}_1,\hat{\i}_2,\hat{\i}_3)$

\begin{equation}
b_r(t)=R_{ra}(t)\,\hat{\i}_a,
\end{equation}

\noindent which is placed on a moving origin, whose inertial
coordinates are $y^a(t)$. The corresponding cartesian coordinates
are the non-inertial (rigid) coordinates $\sigma^r$. This class of
reference frames is obtained in our approach with the choice

\begin{equation}
{\cal A}^a_F(t,\vec{\sigma})=y^a(t)+\sigma^r\,R_{ra}(t).
\label{rigid}
\end{equation}

The time dependent rotation $R(t)$ can be expressed in terms of
time dependent Euler's angles $\alpha(t)$, $\beta(t)$,
$\gamma(t)$. Following the convention of Ref.\cite{LandauQ} the
explicit form of the rotation is

\begin{eqnarray*}
&&R(\alpha,\beta,\gamma)=\\
&&\nonumber\\
&=&\left(
\begin{array}{ccc}
\cos \alpha \cos \beta \cos \gamma -\sin \alpha \sin \gamma  & \sin \alpha
\cos \beta \cos \gamma +\cos \alpha \sin \gamma  & -\sin \beta \cos \gamma
\\
-\cos \alpha \cos \beta \sin \gamma -\sin \alpha \cos \gamma  & -\sin \alpha
\cos \beta \sin \gamma +\cos \alpha \cos \gamma  & \sin \beta \sin \gamma
\\
\cos \alpha \sin \beta  & \sin \alpha \sin \beta  & \cos \beta
\end{array}
\right).
\end{eqnarray*}

Since we have

\begin{equation}
\Omega_{rs}(t)=\left(\frac{d
R(t)}{dt}\,R^T(t)\,\right)_{rs}=-\Omega_{sr}(t),
 \label{V18}
\end{equation}

\noindent we can define the angular velocity

\begin{equation}
\omega^s(t)=\,\frac{1}{2}\varepsilon^{rsu}\Omega_{su}(t).
\end{equation}

\noindent Moreover by defining

\begin{equation}
v^r(t)=R_{ra}(t)\,\frac{dy^a(t)}{dt},
\end{equation}

\noindent we obtain

\begin{equation}
V^r_F(t ,\vec \sigma ) = v^r(t)+
 \Big[\,\vec{\omega}(t)\times\vec{\sigma}\,\Big]^r.
\label{V-rigid}
\end{equation}

\subsection{Rigid Non-Inertial Frames: the Classical Case}

Using the previous results in classical Hamiltonian
(\ref{H-classica-ni}) and defining the components of the {\em
total momentum} and of the {\em total angular momentum} along the
axis of the {\em rotating frames}

\begin{equation}
P^r(t)=\sum_{i=1}^N p^r_i(t ),\qquad
J^r(t)=\sum_{i=1}^N\Big( {\vec \eta}_i(t )\
\times {\vec p}_i(t )\Big)^r,
\end{equation}

\noindent we obtain the classical non-inertial Hamiltonian for
rigid non-inertial frames

\begin{equation}
H_{ni}(t)=\sum_i\,\frac{\vec{p}^2_i(t)}{2m_i}+{\bf V}- \vec{v}(t
)\cdot {\vec P}(t ) - \vec \omega (t ) \cdot \vec{J}(t).
\label{H-classica-rigid}
\end{equation}

\noindent It can be shown that the Hamilton equations obtained using the
Hamiltonian (\ref{H-classica-rigid}) imply

\begin{eqnarray}
 \frac{d^2\vec{\eta}_i(t)}{dt^2}&\on&
-\left( \frac{d\vec{v}(t)}{dt}+
\vec{\omega}(t)\times\vec{v}(t)\right)-
\frac{d\vec{\omega}(t)}{dt}\times\vec{\eta}_i(t) +
\nonumber\\
 &&\nonumber\\
  &-&
2\vec{\omega}(t)\times  \frac{d\vec{\eta}_i(t)}{dt}- \Big(
\vec{\omega}(t)\times (\vec{\omega}(t)\times\vec{\eta}_i(t)) \Big)
-\frac{1}{m_i}\nabla_{\eta_i}{\bf V}.
 \end{eqnarray}

\noindent These are the standard equations of motion of particles in a
non-inertial  rigid reference frame: the four terms in the second
member of the second equation are the {\it standard Euler, Jacobi,
Coriolis and centrifugal forces}, respectively.

\subsection{Rigid Non-Inertial Frames: the Quantum Case}

Using the results (\ref{V-rigid}) in the quantum non-inertial
Hamiltonian (\ref{H-quantistica-ni}), we obtain the {\em quantum
Hamiltonian for rigid non-inertial frames}

\begin{equation}
 \widehat{H}_{ni}(t)=\widehat{E}[{\cal A}_F^a]
+i\hbar\left[ \vec{v}(t)\cdot \sum_i\,\nabla_{\eta_i}\;+\vec{\omega}(t)\cdot
\sum_i\left(\vec{\eta}_i\times\nabla_{\eta_i}\right)
\right].
\label{H-quantistica-rigid}
\end{equation}

\noindent In this rigid case it is useful to show that the explicit form of
$\widehat{E}[{\cal A}_F^a]$ is

\begin{equation}
\widehat{E}[{\cal A}_F^a]=\sum_{i,r}\frac{\hbar^2}{2m_i}\left(
\frac{\partial}{\partial\eta^r_i} \right)+{\bf V}\Big(t,{\cal
A}_F^a(t,\vec{\eta}_1),.... ,{\cal A}_F^a(t,\vec{\eta}_N)\Big).
\end{equation}

Each solution $\psi_F(t,\vec{\eta}_i)$ of the non-inertial
Schroedinger equation with Hamiltonian (\ref{H-quantistica-rigid})
can be obtained by using Eq.(\ref{rigid}) in the general solution
(\ref{trasf-unit-ni-in}). If we  observe that in this case we have
$\det J(t,\vec{\sigma})=\det R(t)=1$, we obtain

\begin{equation}
\psi_F(t,\vec{\eta}_i)=\psi_{in}\Big(t,y^a(t)+\eta_i^r\,R_{ra}(t)\Big)=
{\cal U}_T(t)\cdot{\cal U}_R(t)\cdot\Psi_{in}(t,\vec{\eta}_i).
\label{sol-rigid}
\end{equation}

\noindent Here we used the following time-dependent translations and
rotations

\begin{eqnarray}
&&{\cal U}_T(t)=\exp\left(\frac{i}{\hbar}\sum_{a,r}
y^a(t)\;\widehat{P}^r\,R_{ra}(t)\right),\nonumber\\
&&\nonumber\\
&&{\cal U}_R(t)
=\exp\left(\,\frac{i}{\hbar}
 \gamma(t)\,\widehat{J}^3\right)\,
 \exp\left(\,\frac{i}{\hbar}\beta(t)\,\widehat{J}^2\right)\,
\exp\left(\,\frac{i}{\hbar}
 \alpha(t)\,\widehat{J}^3\right),
 \label{V40}
\end{eqnarray}

\noindent where

\begin{equation}
\widehat{P}^r=-i\hbar\sum_i\,\frac{\partial}{\partial\eta^r_i},\qquad
\widehat{J}^r=-i\hbar\left(\vec{\eta}_i\times\nabla_{\eta_i}\right)^r,
\end{equation}
are the quantum {\em total momentum} and {\em total angular momentum}.

\subsection{Arbitrary Phase Factor and Comparison with other Approaches}

In the non-inertial wave function we can add a arbitrary phase
factor. In other terms a non-inertial observer can choose to
represent a physical state with a wave function

\begin{equation}
\psi^{\,\prime}_F(t,\vec{\eta}_i)=\exp\left(\frac{i}{\hbar}
\Lambda(t,\vec{\eta}_i)\right)\,\psi_F(t,\vec{\eta}_i).
\label{psi-primo2}
\end{equation}

\noindent At the classical level this correspond to a time-dependent
canonical transformation

 \beq
p_{ir}\longmapsto\,p^{\;\prime}_{ir}=p_{ir}+
\frac{\partial\Lambda(t,\vec{\eta}_i)}{\partial\eta^r_i}.
 \eeq

\bigskip

This freedom is useful especially in the rigid case (\ref{rigid}).
As discussed in Appendix A of Ref.\cite{Greenberger}, in this case
a convenient choice is ($\vec{X}=\sum_i\,m_i\vec{\eta}_i/M$,
$M=\sum_i\,m_i$)

\begin{equation}
\frac{i}{\hbar}\Lambda(t,\vec{\eta}_i)=-\frac{i}{\hbar}\,M\,\vec{X}\cdot\vec{v}(t)+F(t).
\label{fase-boost}
\end{equation}

\noindent To add the phase factor is equivalent to do a new time-dependent
unitary transformation. For example the phase (\ref{fase-boost})
is equivalent to the Galileo boost

\begin{equation}
{\cal
U}_B(t,F)=\exp\left(-\frac{i}{\hbar}\,M\,\vec{X}\cdot\vec{v}(t)+F(t)\right).
\end{equation}

\noindent At the classical level, the corresponding canonical transformation
maps the momenta $p_{ir}$'s in  boosted ones

\begin{equation}
p^{\,\prime}_{ir}=p_{ir}-m_i\,v^r(t).
\end{equation}

\noindent In the rigid case (\ref{rigid}), the wave function
(\ref{psi-primo2}), with the choice (\ref{fase-boost}),satisfies
the Schroedinger equation with Hamiltonian

\begin{equation}
 \widehat{H}^{\,\prime}_{ni}(t)=\widehat{E}[{\cal A}_F^a]
+\vec{a}(t)\cdot
\sum_i\,m_i\vec{\eta}_i\;+i\hbar\vec{\omega}(t)\cdot
\sum_i\,\vec{\eta}_i\times\nabla_{\eta_i}
+\left(i\hbar\dot{F}(t)-\frac{1}{2}M\vec{v}^2(t)\right),
\label{H2-quantistica-rigid}
\end{equation}

\noindent where now the acceleration

\begin{equation}
a^r(t)=R_{ra}(t)\,\frac{d^2y^a(t)}{dt^2},
\end{equation}

\noindent gives a momentum-independent potential $+\vec{a}(t)\cdot
\sum_i\,m_i\vec{\eta}_i$. Using Eqs.(\ref{sol-rigid}), a solution
for $\psi_F^{\,\prime}$ can be expressed in the form

\begin{equation}
\psi_F^{\,\prime}(t,\vec{\eta}_i)=e^{i\Lambda(t,\vec{\eta}_i)}\,
{\cal U}_T(t)\cdot{\cal U}_R(t)\cdot\psi_{in}(t,\vec{\eta}_i)\byd
\widetilde{\cal U}_T[y^a(t),F]\cdot{\cal
U}_R(t)\cdot\psi_{in}(t,\vec{\eta}_i),
 \label{klink}
\end{equation}

\noindent where we have defined

 \beq
  \widetilde{\cal
U}_T[y^a(t),F]\byd\,e^{i\Lambda(t,\vec{\eta}_i)}\,{\cal U}_T(t).
\eeq

If  in Eq.(\ref{fase-boost}) we choose
$F(t)=M\vec{y}(t)\cdot\dot{\vec{y}}(t)$, the result (\ref{klink})
reproduces the approach of Ref.\cite{Klink}. In fact the choice
done for $F$ is such that the $\widetilde{\cal U}_T[y^a(t),F]$'s
give the projective representation of the {\em acceleration group}
used in Ref.\cite{Klink}

 \beq
  \widetilde{\cal
U}_T[y^a_1(t),F]\cdot\widetilde{\cal U}_T[y^a_2(t),F]=
\exp\left(\,
-\frac{i}{\hbar}M\dot{\vec{y}}_1(t)\cdot\vec{y}_2(t)\, \right)
\widetilde{\cal U}_T[y^a(t)_1+y^a_2(t),F].
 \eeq

Instead in Ref.\cite{Plebanski} the sequence of time-dependent
unitary transformations ${\cal U}_B(t)\cdot{\cal U}_T(t)\cdot{\cal
U}_R(t)$ is used to construct quantum mechanics in non-inertial
(rigid) frames in the {\em Heisenberg picture}. It can be shown
that the result (\ref{klink}) is equivalent to the approach of
Ref.\cite{Plebanski} with a mapping from the {\em Heisenberg
picture} to the {\em Schroedinger picture}.

\newpage

\section{Center of Mass, Relative Variables and Bound States}

In this Section we discuss the case of two particles mutually
interacting with time-independent, rotationally invariant,
potential, whose form in an inertial frame is

\begin{equation}
{\bf V}={\bf V}\Big( \mid\vec{x}_1-\vec{x}_2\mid \Big).
\label{potenziale-interazione}
\end{equation}

\subsection{Center of Mass and Relative Variables}

Using Eq.(\ref{rigid}), the interaction potential
(\ref{potenziale-interazione}) can be expressed in a rigid
non-inertial frame in the  simple form

\begin{equation}
{\bf V}= {\bf V}\Big( \mid\vec{\cal A}_F(t,\vec{\eta}_1)-\vec{\cal
A}_F(t,\vec{\eta}_2)\mid \Big)= {\bf V}\Big(
\mid\vec{\eta}_1-\vec{\eta}_2\mid \Big).
\end{equation}

\bigskip

In the rigid non-inertial frame defined by Eq.(\ref{rigid}) we can
use a standard separation between center of mass and relative
coordinates

\begin{equation}
X^r=\frac{m_1}{M}\,\eta^r_1+\frac{m_2}{M}\,\eta^r_2,\;\;\;\;\;
\varrho^r=\eta^r_1-\eta^r_2,
\end{equation}

\noindent where we introduced the {\em total mass} $M=m_1+m_2$.
The {\em reduced mass} is $\mu=m_1\,m_2/M$. With standard methods
we can obtain the following expressions

\begin{eqnarray}
&&\widehat{P}^r=-i\hbar\left(
\frac{\partial}{\partial\eta^r_1}+\frac{\partial}{\partial\eta^r_2}
\right)=-i\hbar\,\frac{\partial}{\partial X^r}=-i\hbar\,\nabla^r_{X},\\
&&\nonumber\\
&&\widehat{J}^r=\widehat{L}^r+\widehat{S}^r,\nonumber\\
&&\nonumber\\
&&\widehat{L}^r=-i\hbar\,(\,\vec{X}\times\nabla_X\,)^r,\;\;\;\;\;
\widehat{S}^r=-i\hbar\,(\,\vec{\varrho}\times\nabla_\varrho\,)^r.
\end{eqnarray}

\noindent The Hamiltonian (\ref{H-quantistica-rigid}) can be written as the
sum of a center of mass Hamiltonian and a relative Hamiltonian

\begin{eqnarray}
\widehat{H}_{ni}&=&\widehat{H}_{ni,cm}+\widehat{H}_{ni,rel},\nonumber\\
&&\nonumber\\
\widehat{H}_{ni,cm}&=&-\frac{\hbar^2}{2M}\nabla_X^2+i\hbar\,\vec{v}(t)\cdot\nabla_X+i\hbar
\vec{\omega}(t)\cdot(\,\vec{X}\times\nabla_X\,),\\
&&\nonumber\\
\widehat{H}_{ni,rel}&=&-\frac{\hbar^2}{2\mu}\nabla_\varrho^2+V(\varrho)+i\hbar\,
\vec{\omega}(t)\cdot(\,\vec{\varrho}\times\nabla_\varrho\,).
\label{relativa}
\end{eqnarray}

\noindent As a consequence, there exist solutions of the non-inertial
Schroedinger equation with Hamiltonian (\ref{H-quantistica-rigid})
factorized in center of mass and relative parts
$\psi_F(t,\vec{\eta}_1,\vec{\eta}_2)=\psi_{F,cm}(t,\vec{X})\,
\psi_{F,rel}(t,\vec{\varrho})$. If we make the same separation in
the inertial energy of Eq.(\ref{energia-in}), we obtain

\begin{eqnarray}
\widehat{E}_{in}&=&\widehat{E}_{in,cm}+\widehat{E}_{in,rel},\nonumber\\
&&\nonumber\\
\widehat{E}_{in,cm}&=&-\frac{\hbar^2}{2M}\nabla_X^2,\\
&&\nonumber\\
\widehat{E}_{in,rel}&=&-\frac{\hbar^2}{2\mu}\nabla_\varrho^2+V(\varrho).
\end{eqnarray}

\noindent The corresponding inertial wave function is
$\psi_{in}(t,\vec{\eta}_1,\vec{\eta}_2)
=\widetilde{\psi}(t,\vec{X},\vec{\rho})=
\psi_{in,cm}(t,\vec{X})\,\psi_{in,rel}(t,\vec{\varrho})$. We can
observe that Eqs.(\ref{sol-rigid}) can be rewritten in the
factorized form

\begin{eqnarray}
\psi_{F,cm}(t,\vec{X})&=&{\cal U}_T(t)\cdot{\cal
U}_L(t)\psi_{in,cm}(t,\vec{X}),\\
&&\nonumber\\
\psi_{F,rel}(t,\vec{\varrho})&=&{\cal
U}_S(t)\psi_{in,rel}(t,\vec{\varrho}), \label{sep}
\end{eqnarray}

\noindent where we used the following definitions

\begin{eqnarray}
{\cal U}_L(t)&=&\exp\left(\,\frac{i}{\hbar}
 \gamma(t)\,\widehat{L}^3\right)\,
 \exp\left(\,\frac{i}{\hbar}\beta(t)\,\widehat{L}^2\right)\,
\exp\left(\,\frac{i}{\hbar}
 \alpha(t)\,\widehat{L}^3\right),\\
&&\nonumber\\
{\cal U}_S(t)&=&\exp\left(\,\frac{i}{\hbar}
 \gamma(t)\,\widehat{S}^3\right)\,
 \exp\left(\,\frac{i}{\hbar}\beta(t)\,\widehat{S}^2\right)\,
\exp\left(\,\frac{i}{\hbar}
 \alpha(t)\,\widehat{S}^3\right).
\end{eqnarray}

We can observe that also the Hamiltonian
(\ref{H2-quantistica-rigid}) can be rewritten as the sum of
center of mass part and relative parts

\begin{equation}
\widehat{H}'_{ni}(t)=\widehat{H'}_{ni,cm}(t)+\widehat{H}_{ni,rel}(t),
\end{equation}

\noindent where

\begin{equation}
\widehat{H'}_{ni,cm}(t)=-\frac{\hbar^2}{2M}\,\nabla_{X}^2+
M\vec{a}\cdot\vec{X}+i\hbar\vec{X}\times\nabla_{X}
+\left(i\hbar\dot{F}(t)-\frac{1}{2}M\vec{v}^2(t)\right),
\end{equation}

\noindent and where $\widehat{H}_{ni,rel}(t)$ is the same of
Eq.(\ref{relativa}).

\bigskip

On the contrary, we cannot obtain the center of mass and relative
variable factorization if we use non-rigid non-inertial
coordinates. At the classical level the best  we can do is to
observe that we can return to coordinates in the inertial frame by
a point canonical transformation

\begin{equation}
\eta^{\,\prime\,a}_i={\cal A}^a_F(t,\vec{\eta}_i),\qquad
p^{\,\prime}_{i\,a}=\widetilde{J}^r{}_{Fa}(t,\vec{\eta}_i) p_{ir},
\label{canonica1}
\end{equation}

\noindent and we can apply center of mass and relative variable
transformation to these inertial coordinates by a second canonical
transformation

 \bea
\vec{\rho}=\vec{\eta}^{\,\prime}_1-\vec{\eta}^{\,\prime}_2
&\qquad& \vec{X}=\frac{m_1}{M}\vec{\eta}^{\,\prime}_1
+\frac{m_2}{M}\vec{\eta}^{\,\prime}_2,\nonumber\\
&&\nonumber\\
\vec{\pi}=\frac{m_2}{M}\vec{p}^{\,\prime}_1
-\frac{m_1}{M}\vec{p}^{\,\prime}_2&\qquad&
\vec{P}=\vec{p}^{\,\prime}_1+\vec{p}^{\,\prime}_2.
\label{canonica2}
 \eea

\noindent The inverse total canonical transformation allows to define
non-inertial non-rigid notion of center of mass and relative
variables.

Contrary to {\bf I}, where  a quantum implementation of this classical
approach is very complex, in this non-relativistic case the
inverse total canonical transformation can be implemented with
simple observations. Indeed the inverse of Eq.(\ref{canonica2}) is
implemented by a coordinates transformation on the inertial wave
function written in terms of the relative and center of mass
coordinates $ \psi_{in}(t,\vec{\eta}_1,\vec{\eta}_2)=
\widetilde{\psi}_{in}\Big(t, \vec{X}(\vec{\eta}_1,\vec{\eta}_2),
\vec{\rho}(\vec{\eta}_1,\vec{\eta}_2)\Big) $. This implies that
the energy opertors have to be rewritten as

 \bea
\widehat{E}_{in,cm}&=&-\frac{\hbar^2}{2M}\sum_{a,i}
\left(\frac{\partial}{\partial\eta^a_1}+
\frac{\partial}{\partial\eta^a_2}\right)^2,\nonumber\\
&&\nonumber\\
\widehat{E}_{in,rel}&=&-\frac{\hbar^2}{2\mu}\sum_{a,i}
\left(\frac{m_2}{M}\frac{\partial}{\partial\eta^a_1}-
\frac{m_1}{M}\frac{\partial}{\partial\eta^a_2}\right)^2+ {\bf
V}\Big( \mid\vec{\eta}^{\,\prime}_1-\vec{\eta}^{\,\prime}_2\mid
\Big).
 \eea

Then we can apply the time dependent transformation ${\cal
U}_F(t)$ implementing the inverse of the canonical transformation
(\ref{canonica1}). In particular we obtain the following form of
the non-inertial Hamiltonian

\begin{eqnarray}
\widehat{H}_{ni}&=&\widehat{E}_{cm}[{\cal A}_F^a]+\widehat{E}_{rel}[{\cal A}_F^a]
-{\cal U}_F(t)\frac {d{\cal U}^+_F(t)}{dt},\nonumber\\
&&\nonumber\\
\widehat{E}_{cm}[{\cal A}_F^a]
&\byd&{\cal U}_F(t)\cdot\widehat{E}_{in,cm}\cdot{\cal U}^+_F(t),\\
&&\nonumber\\
\widehat{E}_{rel}[{\cal A}_F^a]&\byd&{\cal U}_F(t)\cdot
\widehat{E}_{in,rel}\cdot{\cal U}^+_F(t), \label{E-rel}
\end{eqnarray}

\noindent where in the {\em quantum inertial potentials} the
center of mass and relative variables remain mixed. As a
consequence the non-inertial wave function cannot be factorized in
general non-rigid non-inertial frames.

\subsection{Bound States}

In inertial frames bounds states are defined looking for
stationary solutions of the Schoredinger equation for the relative
wave function

\begin{equation}
i\hbar\,\frac{d}{dt}\,\psi_{in,rel}(t,\vec{\varrho})=\widehat{E}_{in,rel}\cdot
\psi_{in,rel}(t,\vec{\varrho}),
 \label{L1}
\end{equation}

\noindent whereas the center of mass wave function is a solution
of the equation

\begin{equation}
i\hbar\,\frac{d}{dt}\,\psi_{in,cm}(t,\vec{X})=\widehat{E}_{in,cm}\cdot
\psi_{in,cm}(t,\vec{X}). \label{L2}
\end{equation}

\noindent The stationary solutions of Eq.(\ref{L1}) have the form

\begin{equation}
\psi^{(n)}_{in,rel}(t,\vec{\varrho})=\exp\left(+\frac{i}{\hbar}\,
B_n\,t\right)\,\phi^{(n)}_{in}(\vec{\varrho}), \label{L3}
\end{equation}

\noindent where the $\phi^{(n)}_{in}(\vec{\varrho})$ are a
complete solution of the eingenvalue problem

\begin{equation}
\widehat{E}_{in,rel}\cdot\phi^{(n)}_{in}(\vec{\varrho})
=-B_n\,\phi^{(n)}_{in}(\vec{\varrho}). \label{L4}
\end{equation}

\noindent If $-B_n$ is an eingenvalue in the discrete spectrum ($B_n>0$),
this solution defines a bound state.

Using Eq.(\ref{trasf-unit-ni-in}), the inertial wave function

\begin{equation}
\psi^{(n)}_{in}(t,\vec{\eta}_1,\vec{\eta}_2)=\psi_{in,cm}(t,\vec{X})
\,\psi^{(n)}_{in,rel}(t,\vec{\varrho}), \label{L5}
\end{equation}

\noindent can be mapped in the corresponding non-inertial wave
function

\begin{equation}
\psi^{(n)}_{F}(t,\vec{\eta}_1,\vec{\eta}_2)={\cal U}_F(t)\cdot
\psi^{(n)}_{in}(t,\vec{\eta}_1,\vec{\eta}_2),
 \label{L6}
\end{equation}

\noindent so that Eq.(\ref{E-rel}) implies

\begin{equation}
\widehat{E}_{rel}[{\cal
A}^a]\cdot\psi^{(n)}_{F}(t,\vec{\eta}_1,\vec{\eta}_2)
=-B_n\,\psi^{(n)}_{F}(t,\vec{\eta}_1,\vec{\eta}_2). \label{L7}
\end{equation}

Then, there exist solutions of the non-inertial Schroedinger
equations that are eingenfunctions of the discrete spectrum of the
operator corresponding to the internal energy. These solutions
correspond to bound states defined in inertial frames and they can
be intepreted naturally as the bound states in every non-inertial
(non-rigid) frame.

The previous observations are valid in every non-rigid
non-inertial frame, but {\em only for rigid non-inertial frames,
the non-inertial wave function can still be factorized in  center
of mass and relative parts}. In fact only in this particular case
we can map the factorized inertial solutions
$\psi_{in,cm}(t,\vec{X})$ and $\phi^{(n)}_{in}(\vec{\varrho})$ in
the non-inertial ones using Eqs.(\ref{sep}). In this case we get

\begin{equation}
\psi^{(n)}_{F,rel}(t,\vec{\varrho})={\cal U}_{S}(t)\cdot
\psi^{(n)}_{in,rel}(t,\vec{\varrho})\;\Rightarrow
\;\widehat{E}_{in,rel}\cdot\psi^{(n)}_{F,rel}(t,\vec{\varrho})
=-B_n\,\psi^{(n)}_{F,rel}(t,\vec{\varrho}).
 \label{L8}
\end{equation}

\bigskip

In rigid frames, where a relative non-inertial Hamiltonian
$\widehat{H}_{ni,rel}(t)$, different from the relative energy
$\widehat{E}_{rel}[{\cal A}^a]$, exists, we could  look for a
non-inertial definition of bound states, independent by the
inertial one. For instance we could  look for a stationary
solution of the non-inertial Schroedinger equation for the
relative non-inertial wave function

\begin{equation}
i\hbar\,\frac{d}{dt}\,\widetilde{\psi}_{F,rel}(t,\vec{\varrho})=\widehat{H}_{ni,rel}(t)\cdot
\widetilde{\psi}_{F,rel}(t,\vec{\varrho}), \label{L10}
\end{equation}

\noindent whereas the center of mass wave function is a solution
of one of the non-inertial Schroedinger equations with one of the
center of mass non-inertial Hamiltonians defined in the previous
Sections. Stationary solutions must have the form

\begin{equation}
\widetilde{\psi}^{(n)}_{F,rel}(t,\vec{\varrho})=
\exp\left(+\frac{i}{\hbar}\int^t\,dt_1\,h_n(t_1)\right)\,
\widetilde{\phi}^{(n)}_{F}(\vec{\varrho}),
 \label{L11}
\end{equation}

\noindent where the $\widetilde{\phi}^{(n)}_{F}(\vec{\varrho})$
are solutions of the eigenvalue problem

\begin{equation}
\widehat{H}_{ni,rel}(t)\cdot\widetilde{\phi}^{(n)}_{F}(\vec{\varrho})
=-h_n(t)\,\widetilde{\phi}^{(n)}_{F}(\vec{\varrho}). \label{L12}
\end{equation}

\noindent Eq.(\ref{L12}) is equivalent to a system of infinite eingenvalues
problems (one for each instant "$t$"). But we can show that, in
general, these eingenvalue problems are not simultaneously
solvable. In fact if we take the eingenvalue problems for two
different times $t_1$ and $t_2$, we have

\begin{equation}
[\widehat{H}_{ni,rel}(t_1),\widehat{H}_{ni,rel}(t_2)]=
-i\hbar\vec{\omega}(t_1)\times\vec{\omega}(t_2)\cdot
\Big(\vec{\varrho}\times\nabla_\varrho\Big)\neq 0. \label{L13}
\end{equation}

\noindent Then we cannot define {\em non-inertial bound states} looking for
stationary solution of the relative Schroedinger equation in
general cases. However these stationary solutions can exist in
some particular cases. The most important is the case where the
rotating frame rotates around a fixed axis $\hat{n}$, with angular
velocity $\vec{\omega}(t)=\omega(t)\hat{n}$. In this case, if we
use the set of maximal obsevables $\widehat{E}_{rel}$,
$\widehat{S}^2$, $\widehat{\vec{S}}\cdot\hat{n}$ to label the
solutions of Eq.(\ref{L8}), they are also solutions of
Eq.(\ref{L12}). In other terms we have
$\widetilde{\psi}^{(n)}_{F,rel}=\psi^{(n)}_{F,rel}$. Therefore we
cannot obtain a really different definition of bound states.

\newpage

\section{Two Examples}

In this Section we re-discuss two known cases where the use of a
non-inertial frame is useful. In both of these cases the
interaction potential has an explicit time-dependence in the
inertial reference frame, so that it is more simple to study the
non-inertial Scroedinger equation in a non-inertial frame where
the interaction potential appears time-independent.

\subsection{Cranking Model}

The {\em Cranking Model} \cite{Inglis}, \cite{Bohr-Mottelson},
\cite{review1}, \cite{review2} is a model used in nuclear physics
to study the properties of rapidly rotating non spherical nuclei.
Following Ref.\cite{review1}, in this model it is assumed that the
motion of each nucleon in a non spherical (non rotating) nucleus
is determined by an average potential $V_M(\vec{x})$ without
spherical symmetries. Rapid rotations are described in a
semiclassical manner introducing an active rotation of the
potential $V_M(\vec{x})$ around an axis $\hat{n}$ that is not a
symmetry axis of the potential. In the following we assume
$\hat{n}=\hat{\i}_3$. The Hamiltonian in the inertial (laboratory)
reference frame of a rotating nucleus described with this model is
then

\begin{equation}
\widehat{E}_{in}=\sum_i\,\left[\,
-\frac{\hbar^2}{2m_i}\sum_a\left(
\frac{\partial}{\partial\eta^a_i} \right)^2
+V_M\Big(R_{ra}(t)\eta^a_i\Big)\,\right], \qquad
R(t)=R(0,0,\omega\,t\,).
 \label{ex1}
\end{equation}

\noindent This means to assume an explicit time-dependent potential in the
inertial frame

\begin{equation}
{\bf V}(t,\vec{x}_1,...\vec{x}_N)=\sum_i
V_M\Big(R_{ra}(t)x^a_i\Big).
\end{equation}

\noindent It is convenient to study the system described by the Hamiltonian
(\ref{ex1}) in a rotating reference frame defined in our notation
by the relation

\begin{equation}
{\cal A}^a_F(t,\vec{\sigma})=\sigma^r\,R_{ra}(t).
 \label{ex2}
\end{equation}

\noindent In this frame we must use the non-inertial Schroedinger equation

\begin{equation}
i\hbar\,\frac{\partial}{\partial t}\,
\psi_F(t,\vec{\eta}_1,...\vec{\eta}_N)=\widehat{H}_{ni}(\omega)\cdot
\psi_F(t,\vec{\eta}_1,...\vec{\eta}_N), \label{ex3}
\end{equation}

\noindent where $\widehat{H}_{ni}(\omega)$ is the time-independent
non-inertial Hamiltonian

\begin{equation}
\widehat{H}_{ni}(\omega)=\sum_i\,\left[\,
-\frac{\hbar^2}{2m_i}\sum_r\left(
\frac{\partial}{\partial\eta^r_i}
\right)^2+V_M\Big(\vec{\eta}_i\Big)\right]-\omega\,\widehat{J}^3,
\label{ex4}
\end{equation}

\noindent where, in accord with Eq.(\ref{potential}), we have

\begin{equation}
{\bf V}\Big(t,\vec{\cal A}_F(t,\vec{\eta}_1), ...,\vec{\cal
A}_F(t,\vec{\eta}_N)\Big)=\sum_i\,V_M \Big(\vec{\eta}_i\Big).
\end{equation}

\noindent As stressed in Ref.\cite{review1}, the useful quantity is the {\em
average value of energy in the inertial laboratory frame}. In our
notation (see the comment on the definition (\ref{energia-ni})),
this means the following evaluation

\begin{equation}
\langle E \rangle= \Big(\psi_{F},\widehat{E}[{\cal A}_F^a]\cdot
\psi_{F}\Big)=\Big(\psi_{F},\widehat{H}_{ni}(\omega)\cdot
\psi_{F}\Big)+ \omega\,\Big(\psi_F,\widehat{J}^3\cdot \psi_F\Big).
\label{ex9}
\end{equation}

\subsection{Non Inertial Effects in Interferometry
with Material Waves}

The experimental results of an accelerated or rotating
interferometer for material waves (usually neutrons) compared with
the results of the same interferometer "fixed" in an inertial
(laboratory) frame can be interpreted with the presence of a phase
shift (see Ref.\cite{Libro} and its references).

Following the suggestion of Ref.\cite{Bonse-Wreblowski} a particle
in the fixed interferometer is described by the Hamiltonian

\begin{equation}
\widehat{E}_{in,fixed}=-\frac{\hbar^2}{2m} \left(
\frac{\partial}{\partial\eta^a} \right)^2+V_I\Big(\vec{\eta}\Big),
\label{ex11}
\end{equation}

\noindent where $V_I$ is the potential defined by the cristal in
the interferometer. Since, in general, $V_I$ has no symmetry, a
particle in the same accelerated or rotating interferometer is
described in the inertial (laboratory) frame by the Hamiltonian

\begin{equation}
\widehat{E}_{in}=-\frac{\hbar^2}{2m} \left(
\frac{\partial}{\partial\eta^a}
\right)^2+V_I\Big(R_{ra}(t)\eta^a-(1/2)a^rt^2\Big).
 \label{ex12}
\end{equation}

\noindent Again we have a time-dependent interaction potential in the
inertial frame

 \beq
  {\bf V}(t,\vec{x})=V_I\Big(R_{ra}(t)x^a-(1/2)a^rt^2\Big).
   \eeq

\noindent Then it is useful to study this system in the non-inertial frame

\begin{equation}
{\cal A}^a(t,\vec{\sigma})=\sigma^r\,R_{ra}(t)+
\frac{a^r}{2}\,R_{ra}(t)\,t^2,
 \label{ex13}
\end{equation}

\noindent where we can use the non-inertial Hamiltonian
(\ref{H2-quantistica-rigid}) with $F(t)=(1/6)\,a^2\,t^3$

\begin{equation}
\widehat{H}^{\,\prime}_{ni}(a,\omega)=
\widehat{E}_{in,fixed}+m\vec{a}\cdot\vec{\eta}+i\hbar\vec{\omega}(t)\cdot
\Big(\vec{\eta}\times\nabla_\eta\Big).
 \label{ex14}
\end{equation}

\noindent The inertial potentials that make the Hamiltonian (\ref{ex14})
different from the Hamiltonian (\ref{ex11}) are used to calculate
the observed phase shift \cite{francesi}.

\newpage

\section{Conclusions.}

The main result of this paper is the definition of a quantization
scheme for non-relativistic particle systems in a sufficiently
general class of non-rigid non-inertial frames, including the
usual non-relativistic rigid ones with constant linear
acceleration and angular velocity.

This quantization scheme includes as particular cases many of the
previous results on quantization in non-inertial frames (usually
limited to {\em rigid cases}). As a consequence also the
phenomenological or experimental results based on these
non-inertial quantum theories can be reformulated in the approach
of this paper.

Moreover an original analysis of the definition of bound states in
non-inertial frames is been presented. It turns out that
non-inertial bound states are characterized by the same quantum
numbers of the inertial ones being eigenstates of the {\em
inertial relative energy}, rewritten in terms of non-inertial
coordinates.

\appendix

\newpage

\section{The Non-Relativistic Theory as Exact Limit
of the Relativistic One.}

In I, using Dirac's approach \cite{Dirac}, where a manifestly
covariant Hamiltonian theory is obtained introducing an {\em
admissible foliation} on Minkowski space-time, we have given a
canonical description of a system of $N$ free relativistic
particles on a foliation of parallel hyper-planes.

To describe here this approach, first we have to introduce a {\em
external} inertial frame in Minkowski space-time whose
pseudo-cartesian coordinates are $z^\mu$'s. In such frame we have
to introduce a tetrad of orthonormal four vectors, parametrized
with a 3-vector $\vec{\beta}$

\begin{eqnarray}
U^\mu(\vec{\beta})&=&\left(\,1;\,
\frac{\beta^i}{\sqrt{1-\vec{\beta}^2}}\,\right),\nonumber\\
&&\nonumber\\
\epsilon^\mu_a(\vec{\beta})&=&\left(\,\frac{\beta^a}{\sqrt{1-\vec{\beta}^2}};\,
\delta^{ia}+\frac{\beta^i\beta^a}{\beta^2}
\left(\frac{1-\sqrt{1-\vec{\beta}^2}}{\sqrt{1-\vec {\beta}^2}}
\right) \right),
\end{eqnarray}

\noindent such that

\begin{equation}
U_\mu(\vec{\beta})\,U^\mu(\vec{\beta})=1,\qquad
U_\mu(\vec{\beta})\,\epsilon^\mu_a(\vec{\beta})=0,\qquad
\epsilon^\mu_a(\vec{\beta})\,\epsilon_{\mu\,b}(\vec{\beta})=\eta_{ab}.
\end{equation}

Then we define a foliation with parallel hyper-planes by
introducing the embeddings

\begin{equation}
z^\mu(\tau,\vec{\sigma})=\theta(\tau)\,U^\mu(\vec{\beta})+
\epsilon^\mu_a(\vec{\beta})\,{\cal A}^a(\tau,\vec{\sigma}).
\label{A1}
\end{equation}

\noindent Each hyperplane is defined at $\tau=constant$ and its points are
identified by the curvilinear coordinates $\sigma^r$, implicitly
defined by the invertible coordinates transformation ${\cal
A}^a(\tau,\vec{\sigma})$. The parameter $\tau$ takes the role of
{\em mathematical time} and the function $\theta(\tau)$ describes
the freedom in the choice of this time.

In {\bf I} it is shown how to construct a canonical theory where a
system of $N$ relativistic particles on the hyperplane (\ref{A1})
is described on a phase space whose canonical pairs are:

i) The particles coordinates $\eta^r_i(\tau)$ on the hyperplane at
$\tau$, such that the particle world-lines are

\begin{equation}
x^\mu_i(\tau)=\theta(\tau)U^\mu(\vec{\beta})+
\epsilon^\mu_a(\vec{\beta})\,{\cal A}^a(\tau,\vec{\eta}_i(\tau)),
\end{equation}

\noindent and their momenta $\kappa_{ir}(\tau)$ such that

\begin{equation}
\{\eta^r_i(\tau),\kappa_{js}(\tau)\}=-\delta_{ij}\delta^r_s.
\end{equation}

ii) The degrees of freedom, parametrizing the hyper-planes,
$\theta(\tau)$, ${\cal A}^a(\tau,\vec{\sigma})$ and their momenta,
$M_U(\tau)$, $\rho_a(\tau,\vec{\sigma})$ such that

\begin{equation}
\{\theta(\tau),M_U(\tau)\}=-1,\qquad \{{\cal
A}^a(\tau,\vec{\sigma}),\rho_b(\tau,\vec{\sigma'})\}=
-\delta^a_b\delta^3(\vec{\sigma}-\vec{\sigma}').
\end{equation}

iii) The momentum-like parameter
\[
k^i=\frac{\beta^i}{\sqrt{1-\vec{\beta}^2}},
\]
and their position-like conjugate canonical coordinate $z^i$, such
that

\begin{equation}
\{z^i,k^j\}=-\delta^{ij}.
\end{equation}

On this phase space the dynamics is given by the Dirac Hamiltonian

\begin{equation}
H_D(\tau)=\mu(\tau)\,H_U(\tau)+\int\,d^3\sigma\,\lambda^a(\tau)\,
{\cal H}_a(\tau,\vec{\sigma}),
 \label{A2}
\end{equation}

\noindent where we have used  Dirac's constraints

\begin{eqnarray}
H_U(\tau)&=&M_U(\tau ) - c\, \sum_{i=1}^N\, \sqrt{m^2_i\, c^2
 + \sum_a\,\widetilde{J}^r_a(\tau,\vec{\eta}_i)\,\kappa_{ir}(\tau )\;
 \widetilde{J}^s_a(\tau,\vec{\eta}_i)\,\kappa_{is}(\tau )} \approx 0,\nonumber\\
&&\nonumber\\
{\cal H}_a(\tau,\vec{\sigma})&=&\rho_{a}(\tau,\vec{\sigma}) -
\sum_{i=1}^N\, \delta^3(\vec \sigma - {\vec
 \eta}_i(\tau ))\,\widetilde{J}^r_a(\tau,\vec{\eta}_i)\, \kappa_{ir}(\tau ) \approx
 0.
\end{eqnarray}

\noindent These constraints tell us that the canonical variables
$\theta(\tau),{\cal A}^a(\tau,\vec{\sigma})$ are gauge variables.
As shown in I and in Ref.\cite{sincronizzazione} also in the
relativistic case they are interpreted in terms of non-inertial
frames.

We can see that the canonical coordinates $k^i,z^i$ are trivially
constant on the equations of motion generated by the Dirac
Hamiltonian (\ref{A2}). In the relativistic approach they are
useful to have a manifestly Lorentz covariant canonical theory on
hyper-planes. Since we want to find the non-relativistic limit,
here we have not interested in manifest Lorentz covariance, and we
can eliminate these canonical variables adding by hand a pair of
second class Dirac constraints, which  enforce these variables to
take a constant value

\begin{eqnarray}
k^i\approx\frac{\beta_o^i}{\sqrt{1-\vec{\beta}_o^2}}
=contant,\nonumber\\
&&\nonumber\\
z^i\approx z^i_o=constant.
\end{eqnarray}

As discussed in {\bf I} this step breaks the manifest covariance.

\hfill

The exact non-relativistic limits $c\rightarrow\infty$ is done by
observing that we have

\begin{equation}
c\,\vec{\beta}_o=\vec{u}+{\cal O}(1/c),
\end{equation}

\noindent so that we get

\begin{eqnarray}
U^\mu&=&\delta^\mu_o+ {1\over c}\, (0; \vec u) +{\cal
O}(1/c^2),\nonumber\\
 &&\nonumber\\
\epsilon^\mu_a&=&\delta^\mu_a+ {1\over c}\,
\delta^{\mu}_o\, u^a + {\cal O}(1/c^2).
\end{eqnarray}

By defining $T(\tau)=\theta(\tau)/c$, we arrive at the following
expansion of the {\em embedding}

\begin{eqnarray}
z^o(\tau,\vec{\sigma})&=&c\,T(\tau)+{\cal O}(1/c),\nonumber\\
 &&\nonumber\\
 z^i(\tau,\vec{\sigma})&=&{\cal A}^{a=i}(\tau,\vec{\sigma})+u^i\,T(\tau) + {\cal
 O}(1/c).
 \label{V1}
\end{eqnarray}

\noindent By re-scaling the first equation with a $c$-factor
[$t(\tau,\vec{\sigma}) = z^o(\tau,\vec{\sigma})/c$] we obtain in
the non-relativistic limit

 \beq
 t(\tau,\vec{\sigma}) = T(\tau),\qquad
 z^i(\tau,\vec{\sigma})={\cal
 A}^{i}(\tau,\vec{\sigma})+u^i\,T(\tau).
   \label{V2}
\eeq

This result has the following interpretation: $T(\tau)$ is the
{\em absolute Newtonian time}, while the
$y^i=z^i(\tau,\vec{\sigma})$'s and the $x^a={\cal
A}^a(\tau,\vec{\sigma})$'s are the Cartesian orthogonal
coordinates of two non-relativistic inertial systems with relative
velocity $\vec{u}=constant$ in the {\em 3-dimensional absolute
Newtonian space}.

Let us define the limit of the other variables. The pairs ${\cal
A} ^a(\tau,\vec{\sigma}),\rho_{a}(\tau,\vec{\sigma})$ and
$\eta^r_i(\tau)$, $\kappa_{ir}(\tau)$ are left unchanged by the
non-relativistic limit, while the pair $\theta(\tau),M_U(\tau)$ is
replaced by the pair $T(\tau),K_U(\tau)$ with

\begin{equation}
K_U(\tau)=c\,M_U(\tau)-\sum_{i=1}^N\,m_i\,c^2,
 \label{V3}
\end{equation}

\noindent such that

\begin{equation}
\{K_U(\tau),T(\tau)\}=1.
 \label{V4}
\end{equation}

The form  of the constraints ${\cal H}_a(\tau,\vec{\sigma})
\approx 0$ remains unchanged in the non-relativistic limit.
Instead the constraint $H_U(\tau) \approx 0$ has the following
expansion

 \bea
 H_U(\tau) =\frac{1}{c}\left[\,K_U(\tau)-\sum_i\,\frac{1}{2m_i}\sum_a
\widetilde{J}^r_a(\tau,\vec{\eta}_i)\,\kappa_{i\,r}(\tau)\;
\widetilde{J}^s_a(\tau,\vec{\eta}_i)\,\kappa_{i\,s}(\tau)\,\right]+{\cal O}(1/c^2)
\approx 0,
 \label{V5}
\eea

\noindent and therefore it is replaced by the non-relativistic
constraint

\begin{equation}
H_{U\,
nr}(\tau)=
K_U(\tau)-\sum_i\,\frac{1}{2m_i}\sum_a
\widetilde{J}^r_a(\tau,\vec{\eta}_i)\,\kappa_{i\,r}(\tau)\;
\widetilde{J}^s_a(\tau,\vec{\eta}_i)\,\kappa_{i\,s}(\tau)
\approx 0.
 \label{V6}
\end{equation}

Moreover, we must make the following expansion of the Dirac
multiplier $\mu (\tau )$ in the Dirac Hamiltonian of Eq.(\ref{A2})

\begin{equation}
\mu(\tau)= c\, \rho(\tau) + {\cal O}(1/c),
 \label{V7}
\end{equation}

\noindent if we want to get consistently

\begin{equation}
\frac{d\theta(\tau)}{d\tau}\on-\mu(\tau),\qquad\Rightarrow \qquad
\frac{dT(\tau)}{d\tau}\on-\rho(\tau).
 \label{V8}
\end{equation}

Therefore the non-relativistic Dirac Hamiltonian becomes

\begin{equation}
H_{D,nr}(\tau)=\rho(\tau)\,H_{U\, nr}(\tau)+\int
d^3\sigma\,\lambda^a(\tau,\vec{\sigma})\, {\cal
H}_a(\tau,\vec{\sigma}).
 \label{A3}
\end{equation}

Finally, if we add the gauge fixing $T(\tau)=\tau=t$ (implying
$\rho(\tau)=-1$) we have to substitute the Dirac Hamiltonian
(\ref{A3}) with the Dirac Hamiltonian (\ref{hamiltoniana_dirac}).
We have only to observe that in Subsection IIIB we have used the
canonical momenta
\[
p_{is}=-\kappa_{is},\qquad\rho^a(\vec{\sigma})=-\rho_a(\vec{\sigma}),
\]
to have a non-relativistic sign conventions.

\newpage

\section{Non Relativistic Spinning Particles}

Following Ref.\cite{Spin}, we describe classical spin with
Grassmann degrees of freedom $\xi^a_i(t)$.  We must add to the
Lagrangian (\ref{lagrangiana}) the spin term

\begin{equation}
L_{spin}(\tau)=i\sum_{i,a} \xi_i^a(t)\,\dot{\xi}^a_i(t).
\end{equation}

The only consequence of the presence of the spin Grassmaniann
degrees of freedom on the Hamiltonian formulation is the presence
of the $\xi_i^a(t)$'s as canonical variables \footnote{The
construction of these Poisson brackets is done by using Dirac
brackets to eliminate second class constraints on the conjugate
momenta of the $\xi^a_i$'s.}

\begin{equation}
\{\xi_i^a(t),\xi_j^b(t)\}=i\delta_{ij}\delta^{ab}.
\end{equation}

\noindent Moreover, the potential ${\bf V}$ can depend on the spin degrees
of freedom. The canonical and Dirac Hamiltonians of Section IIIB
are formally unchanged. We must observe that in our construction
the $\xi^a_i(t)$'s represent the components of the spin of the
$i$-th particle along the fixed axes of the inertial frame
$(\hat{\i}_1,\hat{\i}_2,\hat{\i}_3)$.

Canonical quantization maps Poisson bracket of Grassmann variables
into anti-commutators. In our case this means to map the Grassmann
variables $\xi^a_i$ to Pauli matrices

\begin{equation}
\xi^a_i(t)\mapsto\,\sqrt{\frac{\hbar}{2}}\sigma^a_i.
\end{equation}

\noindent Now the wave functions live in the tensor product space
of two-component spinors.

\bigskip

On a fixed non-inertial (non-rigid) frame, the components of the
average value of the spin on the fixed axis
$(\hat{\i}_1,\hat{\i}_2,\hat{\i}_3)$ of the inertial frame is
given by

\begin{eqnarray}
\langle\,s^a_i\,\rangle=\frac{1}{2}(\psi_F,\sigma^a_i\cdot\psi_F).
\end{eqnarray}

If we choose a rigid non-inertial frame, we can also  project the
spin of each particles along the axis of the rotating frame
$(\hat{b}_1(t),\hat{b}_2(t),\hat{b}_3(t))$. To study this
situation, we define first the time-dependent unitary operator

\begin{equation}
{\cal
U}_{R,spin}(t)=\bigotimes_i\,\exp\left(\frac{i}{2}\gamma(t)\sigma_i^3\right)
\,\exp\left(\frac{i}{2}\beta(t)\sigma_i^2\right)
\,\exp\left(\frac{i}{2}\alpha(t)\sigma_i^3\right),
\end{equation}

\noindent and the wave function

\begin{equation}
\psi^{\;\prime}_F(t,\vec{\eta}_i)={\cal
U}_{R,spin}(t)\cdot\psi_F(t,\vec{\eta}_i).
\end{equation}

\noindent Then the components of the average value of the spin along the
rotating axis $(\hat{b}_1(t),\hat{b}_2(t),\hat{b}_3(t))$ are

\begin{eqnarray}
\langle\,\overline{s}^r_i\,\rangle=\frac{1}{2}\,R_{ra}(t)(\psi_F,\sigma^a_i\cdot\psi_F)
=\frac{1}{2}(\psi^{\;\prime}_F,\sigma^r_i\cdot\psi^{\;\prime}_F).
\end{eqnarray}

\noindent This means that we can represent the components of the spin of the
i-th particles along the axis of the rotating frame with the
operator

\begin{equation}
\overline{s}^r_i=\frac{\sigma^r_i}{2},
\end{equation}

\noindent only if we use the wave function
$\psi^{\;\prime}_F(t,\vec{\eta}_i)$. Since the wave function
$\psi_F(t,\vec{\eta}_i)$ satisfies the non-inertial Schroedinger
equation with Hamiltonian (\ref{H-quantistica-rigid}) and since we
have

\begin{equation}
{\cal U}_{R,spin}(t)\frac{d{\cal U}^+_{R,spin}(t)}{dt}=
\sum_i\,\vec{\omega}(t)\cdot\vec{\sigma}_i \;{\cal U}_{R,spin}(t),
\end{equation}

\noindent we get

\begin{equation}
i\hbar\frac{\partial}{\partial t}\psi^{\;\prime}_F(t,\vec{\eta}_i)
= \widehat{H}^s_{ni}(t)\cdot\psi^{\;\prime}_F(t,\vec{\eta}_i),
\end{equation}

\noindent with

\begin{equation}
 \widehat{H}^s_{ni}(t)=\widehat{E}_{in}[{\cal A}^a]
+i\hbar\left[ \vec{v}(t)\cdot \sum_i\,\nabla_{\eta_i}\;+\vec{\omega}(t)\cdot
\sum_i\left(\,\vec{\eta}_i\times\nabla_{\eta_i}+2\vec{\sigma}_i\,\right)
\right].
\end{equation}

\bigskip

The presence of the term $\vec{\omega}\cdot\vec{\sigma}_i$ is
discussed for experimental tests in Ref.\cite{mashoon}

\newpage

\section{Galilei Transformations}

In Subsection IIA we started choosing an inertial frame with
inertial coordinates $x^a$ in Newtonian space-time. However we
could  have started with another inertial frame whose inertial
coordinates $x^{\,\prime\,a}$ are obtained by means of a Galilei
transformation

\begin{equation}
x^{\,\prime\,a}=R^a_b\,x^b+v^a\,t+b^a.
\end{equation}

At the Lagrangian level of Subsection IIB this implies the
following Lagrangian transformations of the functions ${\vec {\cal
A}}(t, \vec \sigma )$

\begin{equation}
{\cal A}^{\,\prime\,a}(t,\vec{\sigma})= R^a_b\,{\cal
A}^b(t,\vec{\sigma})+v^a\,t+b^a.
 \label{galileo1}
\end{equation}

At the  Hamiltonian level  the point transformation
(\ref{galileo1}) is completed with the transformation properties
of the canonical momenta

\begin{eqnarray}
{\rho}^{\,\prime\,a}(t,\vec{\sigma})&=&
R^a_b\,{\rho}^b(t,\vec{\sigma})+\sum_i\,m_i\,v^a
\delta^3(\vec{\sigma}-\vec{\eta}_i),\nonumber\\
&&\nonumber\\
p^{\,\prime}_{ir}(t)&=&p_{ir}+J_r^{\,\prime}{}^a(t,\vec{\eta}_i)\,v_a.
\label{galileo2}
\end{eqnarray}

\noindent Eqs.(\ref{galileo1}) and (\ref{galileo2}) define a time dependent
canonical transformation that leaves unchanged the structure of
the Dirac Hamiltonian since we have ($\lambda^{\,\prime\,a} =
\lambda^a+v^a$)

\begin{equation}
H_D\mapsto H_D^{\,\prime}=H_c^{\,\prime}+\int d^3\sigma\,
\lambda^{\,\prime\,a}(t,\vec{\sigma})\,{\cal
H}_a^{\,\prime}(t,\vec{\sigma})-
\frac{1}{2}\sum_i\,m_i\vec{v}^2(t),
\end{equation}

\noindent where $H_c^{\,\prime}$ and ${\cal
H}_a^{\,\prime}(t,\vec{\sigma})$ are the canonical Hamiltonian and
the constraints expressed in terms of ${\cal
A}^{\,\prime\,a}(t,\vec{\sigma})$,
${\rho}^{\,\prime\,a}(t,\vec{\sigma})$, $p^{\,\prime}_r(t)$ by
inversions of Eqs. (\ref{galileo1})(\ref{galileo2}) and where
$\frac{1}{2}\sum_i\,m_i\vec{v}^2(t)$ is a ignorable function only
of time. In particular, the form of the constraints is left
unchanged by the canonical transformation

\beq
{\cal H}_a^{\,\prime}(t,\vec{\sigma})=
\rho^{\,\prime\,a}(t,\vec{\sigma})-
\sum_i\delta^3(\vec{\sigma}-\eta_i(t))\,
\widetilde{J}^{\,\prime\,r}{}_a(t,\vec{\eta}_i(t))\,p^{\,\prime}_{ir}(t)
\approx 0.
\eeq

\noindent Instead the canonical Hamiltonian becomes

\begin{eqnarray}
H_c(t)
&=&\sum_i\,\frac{1}{2m_i}\sum_a\,
\Big[\widetilde{J}^{\,\prime\,r}{}_a(t,\vec{\eta}_i(t))\,
p_{ir}(t)\Big]
\Big[\widetilde{J}^{\,\prime\,s}{}_a(t,\vec{\eta}_i(t))\,
p_{is}(t)\Big]+\nonumber\\
&&\nonumber\\
&+& \widetilde{\bf V}\Big(t,\vec{\cal
A}^{\,\prime}(t,\vec{\eta}_1),..., \vec{\cal
A}^{\,\prime}(t,\vec{\eta}_N)\Big),
\end{eqnarray}

\noindent with

\begin{equation}
\widetilde{\bf V}\Big(t,\vec{\cal
A}^{\,\prime}(t,\vec{\eta}_1),..., \vec{\cal
A}^{\,\prime}(t,\vec{\eta}_N)\Big)= {\bf V}\Big(t,\vec{\cal
A}(t,\vec{\eta}_1),..., \vec{\cal A}(t,\vec{\eta}_N)\Big).
\end{equation}

\bigskip

In the rest of this Appendix we assume that the interaction
potential ${\bf V}(t,\vec{x}_1,...,\vec{x}_N)$ is time-independent
and invariant under rotations and translations, namely that we
have

\begin{equation}
\widetilde{\bf V}\Big(\vec{\cal A}^{\,\prime}(t,\vec{\eta}_1),...,
\vec{\cal A}^{\,\prime}(t,\vec{\eta}_N)\Big)= {\bf
V}\Big(\vec{\cal A}^{\,\prime}(t,\vec{\eta}_1),..., \vec{\cal
A}^{\,\prime}(t,\vec{\eta}_N)\Big).
\end{equation}

\noindent Then the form of the canonical Hamiltonian is left
unchanged by the Galileo canonical transformation. In this way
Galilei relativity principle is implemented in our parametrized
Galilei theory.

The canonical generators of the transformation (\ref{galileo1})
and (\ref{galileo2}) are (the Galilei boosts are ${\vec {\cal K}}
-+t\, {\vec {\cal P}}$)

\begin{eqnarray}
\vec{\cal J}(t)&=&\int d^3\sigma\,
\vec{\cal A}(t,\vec{\sigma})\times\vec{\rho}(t,\vec{\sigma}),\label{galileo3a}\\
&&\nonumber\\
\vec{\cal P}(t)&=&\int d^3\sigma\,\vec{\rho}(t,\vec{\sigma}),\label{galileo3b}\\
&&\nonumber\\
\vec{\cal K}(t)&=& -\sum_i\,m_i\,\vec{\cal A}(t,\vec{\eta}_i),
\label{galileo3c}
\end{eqnarray}

\noindent and an infinitesimal Galilei transformation is given by

\begin{equation}
\delta F=\{F,G\},\;\;\mbox{ with }\;\;
G=\delta\vec{\omega}\cdot\vec{\cal
J}(t)+\delta\vec{b}\cdot\vec{\cal P}(t)+
\delta\vec{v}\cdot\Big(\vec{\cal K}(t)+t\,\vec{\cal P}(t)\Big).\\
\end{equation}

When $\{{\bf V},\vec{\cal J}(t)\}=\{{\bf V},\vec{\cal P}(t)\}=0$
we get $\{H_c,\vec{\cal J}(t)\}=\{H_c,\vec{\cal P}(t)\}=0$. Then
Eqs. (\ref{galileo3a}), (\ref{galileo3b}) and (\ref{galileo3c})
and the hamiltonian $H_c$ are the generators of a realization of
the Galilei Lie algebra on phase space.

\hfill

At the quantum level  the rules of Subsection IIIA would map the
momenta $\rho^a(t,\vec{\sigma})$, appearing in the infinitesimal
generators (\ref{galileo3a}) and (\ref{galileo3b}),  in the
functional derivatives $\delta/\delta{\cal A}^a(\vec{\sigma})$. As
noted in  I for the canonical generators of the Poincar\'e group,
these functional derivatives are not operators in the Hilbert
space ${\bf H}$, so that  we would not obtain a representation of
the Galileo algebra on the Hilbert space. However, since we are
interested only in the transformation properties of the physical
states, solutions of the generalized Schroedinger equations, we
can substitute the functional derivative with the generalized
hamiltonian $\widehat{T}(\vec{\sigma},{\cal A}^a]$. In this way we
have (see Eq.(3.8) for the definition of $\widehat{K}_{ia} =
\widehat{K}_i^a$)

\begin{eqnarray}
\widehat{\cal J}^a(t)&=&\sum_i\,\epsilon^{abc}
{\cal A}^b(t,\vec{\eta}_i)\,\widehat{K}^c_i,\label{galileo4a}\\
&&\nonumber\\
\widehat{\cal P}^a(t)&=&\sum_i \widehat{K}^a_i,\label{galileo4b}\\
&&\nonumber\\
\widehat{\cal K}^a(t)&=& -\sum_i\,m_i\,{\cal A}^a(t,\vec{\eta}_i).
\label{galileo4c}
\end{eqnarray}

Now these are self adjoint operators. When the interaction
potential is invariant under rotations and translations, that is
when $[{\bf V},\widehat{\cal J}^a(t)]=[{\bf V},\widehat{\cal
P}^a(t)]=0$ implies $[\widehat{E}[{\cal A}^a],\widehat{\cal
J}^a(t)]=[\widehat{E}[{\cal A}^a],\widehat{\cal P}^a(t)]=0$, Eqs.
(\ref{galileo4a}), (\ref{galileo4b}), (\ref{galileo4c}) and the
energy $\widehat{E}[{\cal A}^a]$ become the generators of a
commutator projective realization the Galilei Lie algebra  on the
Hilbert space {\bf H}.

\end{document}